\documentclass[11pt,a4paper]{article}

\usepackage{amsmath,amsfonts,amsbsy,amssymb,array,accents,dsfont}
\usepackage{enumerate,array,latexsym,graphicx,mathrsfs,verbatim,psfrag}
\usepackage{bm} 
\usepackage{bbm,braket}
\usepackage[normalem]{ulem}
\usepackage{booktabs} 
\usepackage[usenames]{color}

\usepackage[UKenglish]{babel}
\usepackage{fancybox}
\usepackage{relsize}

\usepackage[nosort]{cite}
\usepackage{chngpage} 

\usepackage{enumitem}
\setlist{itemsep=0pt}

\usepackage[colorlinks=true,      linkcolor=darkblue,      urlcolor=darkblue,      
            filecolor=darkblue,      citecolor=darkblue,       pdfstartview=FitH,     
						pdfpagemode=UseNone,      bookmarksopen=true]{hyperref}  
\usepackage[all]{hypcap}     

\newcommand{\captionfonts}{\small}
\makeatletter  
\long\def\@makecaption#1#2{%
  \vskip\abovecaptionskip
  \sbox\@tempboxa{{\captionfonts #1: #2}}%
 \ifdim \wd\@tempboxa >\hsize
    {\captionfonts #1: #2\par}
  \else
    \hbox to\hsize{\hfil\box\@tempboxa\hfil}%
  \fi
  \vskip\belowcaptionskip}
\makeatother   

\DeclareMathSymbol{\medhatsym}{\mathord}{largesymbols}{"62} 

\DeclareMathSymbol{\medtildesym}{\mathord}{largesymbols}{"65}
\newcommand\lowermedtildesym{
  \text{\smash{\raisebox{-1.35ex}{%
    $\medtildesym$}}}}
\newcommand\medtilde[1]{
  \mathchoice
    {\accentset{\displaystyle\lowermedtildesym}{#1}}
    {\accentset{\textstyle\lowermedtildesym}{#1}}
    {\accentset{\scriptstyle\lowermedtildesym}{#1}}
    {\accentset{\scriptscriptstyle\lowermedtildesym}{#1}}
}

\newcommand\scalemath[2]{\scalebox{#1}{\mbox{\ensuremath{\displaystyle #2}}}}

\def\({\left(}
\def\){\right)}
\def\[{\left[}
\def\]{\right]}

\def\barray{\begin{array}}
\def\earray{\end{array}}
\def\be{\begin{equation}}
\def\ee{\end{equation}}
\def\bea{\begin{eqnarray}}
\def\eea{\end{eqnarray}}
\def\bal{\begin{align}}
\def\eal{\end{align}}

\numberwithin{equation}{section} %

\makeatletter
\g@addto@macro\bfseries{\boldmath}
\makeatother

\definecolor{cardinal}{rgb}{0.6,0,0}
\definecolor{darkgreen}{rgb}{0,0.4,0}
\definecolor{purple}{rgb}{0.5, 0, 0.5}
\definecolor{golden}{rgb}{0.92, 0.7, 0}
\definecolor{midnight}{rgb}{0, 0, 0.5}
\definecolor{darkblue}{rgb}{0, 0, 0.8}

\definecolor{emeraude}{RGB}{34,120,15}
\definecolor{turquoise}{RGB}{49, 140, 231}
\definecolor{framboise}{RGB}{199, 44, 72}

\def\cC{{\cal C}}
\def\cD{{\cal D}}

\def\cG{{\cal G}}
\def\cH{{\cal H}}
\def\cJ{{\cal J}}
\def\cK{{\cal K}}

\def\cM{{\cal M}}
\def\cN{{\cal N}}
\def\cO{{\cal O}}

\def\sst#1{\scriptscriptstyle{#1}}

\def\tr{{\rm Tr}}

\newcommand{\alphaone}{\alpha_1}
\newcommand{\alphatwo}{\alpha_2}

\begin{document}

\begin{flushright}
%
%
\end{flushright}

\vspace{15mm}

\begin{center}

{\huge \bf{Holographic correlators from multi-mode 
\vspace{5mm}\\
AdS$_5$ bubbling geometries}} \\[5mm]

\vspace{20mm}

{\large
\textsc{David Turton,\;\,Alexander Tyukov}}

\vspace{15mm}

\baselineskip=16pt
\parskip=3pt

Mathematical Sciences and STAG Research Centre,\\ University of Southampton,\\ Highfield,
Southampton SO17 1BJ, UK \\

\vspace{12mm}

{\upshape\ttfamily d.j.turton @ soton.ac.uk, a.tyukov @ soton.ac.uk} \\

\vspace{17mm}

\textsc{Abstract}

\end{center}

\begin{adjustwidth}{9.5mm}{10.5mm} 
%
\vspace{3mm}
\baselineskip=14.5pt
\noindent
Supergravity calculations of holographic four-point correlation functions are a powerful tool for deriving dynamical information about the dual field theory at strong coupling.
Recently, a number of such computations have been performed by studying light probes of smooth horizonless supergravity solutions, including certain asymptotically AdS$_5 \times$S$^5$ solutions of Lin-Lunin-Maldacena (LLM) type. 
We construct a new closed-form perturbative LLM solution involving a pair of linearized supergravitons with different mode numbers and their quadratic backreaction.
Using this solution, we compute two infinite sequences of four-point correlators.
Depending on the scaling of certain parameters, the background supergravity solutions are dual to states of 4D $\mathcal{N}=4$ super Yang-Mills theory that are either heavy or light, whereupon the correlators in the dual field theory are either (perturbatively) heavy-heavy-light-light or all-light.
In the all-light regime, the correlators we compute are related by superconformal Ward identities to correlators of four single-particle chiral primary operators (CPOs). 
We confirm a set of expressions in the literature for such correlators of CPOs, that were obtained via Mellin space bootstrap and/or Witten diagram methods.
We also obtain a new explicit compact expression for one of the two infinite sequences of correlators of chiral primaries.
The method we present is capable of accessing correlators with an arbitrary degree of extremality.

\end{adjustwidth}

\thispagestyle{empty}

\newpage


%
%


\baselineskip=16pt
\parskip=3pt


\section{Introduction}

Holographic duality is the striking and now familiar concept that a gravitational theory can be dual to a non-gravitational theory defined in fewer spacetime dimensions.
The prototypical example is the conjectured duality between type IIB string theory on five-dimensional Anti-de Sitter space times a five-sphere, AdS$_5 \times$S$^5$, and four-dimensional $\cN=4$ SU($N$) super Yang-Mills (SYM) theory on the conformal boundary of AdS$_5$~\cite{Maldacena:1997re,Gubser:1998bc,Witten:1998qj}.
The supergravity limit in AdS$_5 \times$S$^5$ corresponds to the regime of large central charge $c$ (implying large $N$) and strong 't Hooft coupling in the dual field theory.

Four-point correlation functions of single-particle supergravity Kaluza-Klein (KK) modes have provided valuable dynamical information about strongly coupled $\cN=4$ SYM.
Over several years, various families of such correlators were computed at tree level using Witten diagrams~\cite{Arutyunov:2000py,Arutyunov:2002fh,Arutyunov:2003ae,Berdichevsky:2007xd,Uruchurtu:2008kp,Uruchurtu:2011wh}. 
These KK modes correspond to chiral primary operators (CPOs) of the SYM theory. For the purpose of computing many observables, the dual operators can be taken to be single-trace CPOs. However, more precisely, the dual operators involve multi-trace admixtures, and are known as single-particle operators~\cite{Aprile:2020uxk}, see also~\cite{Arutyunov:1999en,Arutyunov:2000ima,Rawash:2021pik}.

Using a Mellin space~\cite{Mack:2009mi,Penedones:2010ue,Fitzpatrick:2011ia} and CFT bootstrap approach, a conjecture for the general family of these correlators was developed in~\cite{Rastelli:2016nze,Rastelli:2017udc}. This conjecture was verified in further examples via Witten diagram calculations~\cite{Arutyunov:2018neq}, including correlators of all single-particle chiral primary operators up to dimension 8, plus a selection of others~\cite{Arutyunov:2018tvn}. Soon after, a generating function for the general family of these correlators was conjectured and verified in many examples~\cite{Caron-Huot:2018kta}, and it was shown to imply the Mellin space proposal of~\cite{Rastelli:2016nze,Rastelli:2017udc}. 
Related interesting observables include integrated four-point correlators, see e.g.~\cite{Binder:2019jwn,Dorigoni:2021bvj,
Brown:2023zbr,Alday:2023pet}, 
as well as correlators (both integrated and unintegrated) in a particular large-charge 't Hooft limit~\cite{Paul:2023rka,Caetano:2023zwe,Brown:2024yvt}.

Recently, heavy-heavy-light-light (HHLL) and all-light (LLLL) four-point correlators in AdS$_5$/CFT$_4$ holography have been computed using light probes of smooth horizonless supergravity solutions~\cite{Turton:2024afd,Aprile:2024lwy,Aprile:2025hlt}. 
We use `heavy' to denote CFT operators whose conformal dimensions scale linearly with $c$ in the large $c$ limit, while `light' denotes operators whose scaling dimensions are order one compared to $c$. 
The supergravity backgrounds that were used are particular asymptotically AdS$_5 \times$S$^5$ Lin-Lunin-Maldacena (LLM) solutions~\cite{Lin:2004nb}.
LLM solutions are specified by a colouring of a two-dimensional plane into black and white regions, arising from a smoothness condition.
In $\cN=4$ SYM, the dual 1/2-BPS states are multi-trace CPOs composed of a holomorphic linear combination of two of the six hermitian scalars, $Z = \phi_1 + i \phi_2$. The conformal weight $\Delta$ of such operators is equal to a U(1) R-charge $J$ in the SO(6) R-symmetry group~\cite{Lin:2004nb}.

The LLM background used to compute correlators in~\cite{Turton:2024afd} is specified by a single black droplet with a boundary that consists of an area-preserving ripple deformation of a circle, that had previously been studied in~\cite{Skenderis:2007yb}. In plane polar coordinates $(r,\tilde\phi)$, it is given by
\be
\label{eq:ripplon-prof-TT1}
	r(\tilde\phi)\,=\, \sqrt{1+\alpha  \cos (n \tilde\phi )}\,,
\ee 
with $n=2$ and $|\alpha|<1$. Using as probes the KK modes of the dilaton/axion, an infinite sequence of dynamical four-point correlators were computed. The correlators consist of two dimension-two chiral primaries $\cO_2=\tr Z^2$ (arising from the background) and two descendants $\cD_k \sim \bar{Q}^4 \cO_{k+2}$ (corresponding to the probe), where $\cO_{k}\sim \tr Z^{k}$ up to multi-trace admixtures, which will not play a role in this paper. The CPO $\cO_k$ has $\Delta=J=k$, while the descendant $\cD_k$ has $J=k$, $\Delta=k+4$.

By contrast, a different LLM solution specified by a single elliptical droplet was used to compute correlators in~\cite{Aprile:2024lwy,Aprile:2025hlt}. This solution has been referred to as a 1/2-BPS ``AdS bubble'' in the literature~\cite{Chong:2004ce,Liu:2007xj,Chen:2007du,Giusto:2024trt}. 
It lies in an SO(4)-invariant consistent truncation derived in~\cite{Cvetic:2000nc}.\footnote{A U(1)$^3$-invariant truncation was also derived in\cite{Cvetic:2000nc}; this contains 1/4- and 1/8-BPS AdS bubble solutions~\cite{Chong:2004ce,Liu:2007rv,Ganchev:2025dzn}.}
For this reason, it was recently proposed that the corresponding dual CFT state is a coherent state composed only of powers of the lightest CPO, $\cO_2$~\cite{Giusto:2024trt}, and this proposal was made precise in~\cite{Aprile:2025hlt}.  
By probing this solution with the zero mode of the dilaton/axion, holographic four-point correlators involving two single-traces $\cO_2$ and two double- or triple-trace operators, $\cO_2^2$ or $\cO_2^3$, were computed in~\cite{Aprile:2024lwy,Aprile:2025hlt}, yielding novel dynamical information about $\cN=4$ SYM.

Using precision holographic three-point functions, the proposal that the dual CFT state to the elliptical droplet solution contains only powers of $\cO_2$ has been confirmed up to quadratic order in $\alpha$~\cite{Turton:2025svk}. In that work, it was also demonstrated that the solution specified by the profile~\eqref{eq:ripplon-prof-TT1} is holographically dual to a CFT state that contains also a specific admixture of $\cO_4$ at order $\alpha^2$.
These developments on AdS$_5$/CFT$_4$ holographic correlators build on previous 
work in AdS$_3$/CFT$_2$ duality on HHLL four-point correlators~\cite{Galliani:2017jlg,Bombini:2017sge,Bombini:2019vnc,Giusto:2023awo} and computing LLLL correlators by making extrapolations thereof~\cite{Giusto:2018ovt,Giusto:2019pxc,Giusto:2020neo,Ceplak:2021wzz}; see also the related works~\cite{Bena:2019azk,Rastelli:2019gtj,Giusto:2020mup,Bufalini:2022wyp,Bufalini:2022wzu}.
Some other recent work involving LLM solutions can be found in~\cite{deMelloKoch:2018ert,
Balasubramanian:2018yjq,
Holguin:2023orq,
Berenstein:2023vtd,
Eleftheriou:2023jxr,
Deddo:2024liu,
Chen:2024oqv,
Anempodistov:2025maj}.

Although the Witten diagram method of~\cite{Arutyunov:2018tvn} can compute four-point correlators of CPOs of any specific weights of manageable size numbers (the highest example given in that work is a weight of 25), the method of probing smooth supergravity solutions has the advantage of yielding closed-form expressions for infinite sequences of correlators in which two of the operators (corresponding to the probe) have arbitrarily large weights. This method is therefore capable of making much more general tests of the CFT conjectures of~\cite{Rastelli:2016nze,Rastelli:2017udc,Caron-Huot:2018kta}. 
However, doing so requires more general closed-form backgrounds with which to compute correlators.

In this paper we compute a new closed-form perturbative solution specified by the two-mode profile function
\be
\label{eq:ripplon-prof-multi-intro}
	r(\tilde\phi)\,=\, \sqrt{1+\alpha_1  \cos (n \tilde\phi )+\alpha_2 \cos (m \tilde\phi )}\,,
\ee 
where $|\alpha_1|+|\alpha_2|<1$, and we take $m>n \ge 2$ without loss of generality. We focus primarily on $n=2$, $m=3$, however one can carry out our method for any specific choice of the pair of integer mode numbers $m,n$, and we give some intermediate results for general $m,n$.

We work to quadratic order in perturbation theory in small $\alpha_1$, $\alpha_2$. At linear order, the background consists of a pair of linearised supergravity fluctuations among those classified in~\cite{Kim:1985ez}, see~\cite{Grant:2005qc}. At order $\alpha_1^2$ and $\alpha_2^2$, the background involves the backreaction of each individual mode separately, the same as for the single-mode profile~\eqref{eq:ripplon-prof-TT1}.
The main novelty is at order $\alpha_1\alpha_2$, where there is a non-trivial interaction between modes that gives rise to terms in the supergravity fields with frequencies $m \pm n$. 
The resulting solution for $n=2$, $m=3$ is the first closed-form multi-mode asymptotically AdS$_5 \times $S$^5$ LLM solution. It is analogous to multi-mode solutions in AdS$_3 \times $S$^3$ of the two-charge~\cite{Lunin:2001jy,Lunin:2002iz,Kanitscheider:2007wq,Bombini:2017sge} or three-charge superstrata~\cite{Bena:2015bea,Bena:2016agb,Bena:2016ypk,Bena:2017xbt,Ceplak:2018pws,Heidmann:2019zws,Heidmann:2019xrd,Mayerson:2020tcl,Ganchev:2021iwy,Ceplak:2022pep,Ceplak:2024dbj} types.

The scaling of the parameters $\alpha_1$, $\alpha_2$ with $c$ determines whether the dual CFT states to these  backgrounds are heavy or light. At large $c$, the energy above the global AdS$_5 \times$S$^5$ vacuum scales as $(\alpha_1^2+\alpha_2^2)c$. 
When $\alpha_1^2+\alpha_2^2$ is of order 1 compared to $c$, the state is heavy, and when also $\alpha_{1},\alpha_{2} \ll 1$, the state can be described as ``perturbatively heavy'', see e.g.~\cite{Balasubramanian:2017fan}. 
By contrast, when $\alpha_1^2+\alpha_2^2$ is taken to scale as $1/c$, the state is light. In our supergravity approach, the distinction between perturbatively heavy and light is not seen; there is a unique correlator. 
In other words, upon expanding the HHLL correlators for small $\alpha_{1},\alpha_{2}$ to quadratic order, we obtain LLLL correlators. For related discussions of this point, see~\cite{Ceplak:2021wzz,Turton:2024afd,Aprile:2025hlt}.

We then compute two infinite sequences of holographic correlators, and analyze them in the all-light limit.
The first sequence, $\langle \mathcal{O}_3 \bar{\mathcal{O}}_3 \bar{\mathcal{D}}_{k}  \mathcal{D}_{k}  \rangle$, requires only the terms in the metric proportional to $\alpha_2^2$, arising from the mode $m=3$ (or equivalently, the solution specified by the single-mode profile \eqref{eq:ripplon-prof-TT1} with $n=3$). 
This sequence is obtained via a direct generalisation of the method of~\cite{Turton:2024afd} from $n=2$ to $n=3$. 
By contrast, the second sequence, 
$\langle \mathcal{O}_2 \bar{\mathcal{O}}_3  \bar{\mathcal{D}}_{k} \mathcal{D}_{k+1} \rangle$,
crucially requires the details of the non-trivial interaction between the two modes $n=2$ and $m=3$, as encoded in the terms proportional to $\alpha_1\alpha_2$ that involve the mode $m-n$.
As a side remark, the terms that involve the mode $m+n$ correspond to a sequence of non-dynamical extremal correlators, so we do not study these terms in detail.

These sequences of correlators of two CPOs and two descendants are related by a superconformal Ward identity to correlators of four single-particle CPOs.
By making a prescription for this Ward identity, we find precise agreement with a set of expressions in the literature for correlators of CPOs~\cite{Rastelli:2016nze,Rastelli:2017udc,Arutyunov:2018neq,Arutyunov:2018tvn,Caron-Huot:2018kta,Aprile:2017xsp}.
Explicitly, the first sequence of correlators is related to 
$\langle \cO_3 \bar\cO_3 \bar\cO_{p} \cO_{p}\rangle$, for which an expression for general $p\ge 3$ was derived in~\cite{Aprile:2017xsp}, using the 
Mellin space formula of~\cite{Rastelli:2016nze,Rastelli:2017udc}. To our knowledge, our calculation is the first confirmation of this expression for general $p$ from a supergravity computation.

The second sequence is related to $\langle \cO_2 \bar\cO_3 \bar\cO_{p} \cO_{p+1}\rangle$, several examples of which were computed via Witten diagrams in~\cite{Arutyunov:2018neq,Arutyunov:2018tvn}, up to $p=7$. 
We first verify that the superconformal Ward identity is satisfied for $\langle \cO_2 \bar\cO_3 \bar\cO_{4} \cO_{5}\rangle$ in precisely the form given in~\cite{Arutyunov:2018neq}. We then derive a compact explicit expression for the infinite sequence $\langle \cO_2 \bar\cO_3 \bar\cO_{p} \cO_{p+1}\rangle$ for any $p$. In the finite range of overlap, $2 \le p \le 7$, this expression agrees with the examples given in~\cite{Arutyunov:2018tvn}. Moreover, we compare this expression with the conjecture of~\cite{Rastelli:2016nze,Rastelli:2017udc} for general $p$, finding precise agreement. 
We also check that the expression is consistent with the proposed generating function of~\cite{Caron-Huot:2018kta}.

We note that by implementing the method of the present work to two modes of general $m$ and $n$, one can access correlators with an arbitrary degree of extremality. This can be seen as follows. Consider a correlator of CPOs of highest/lowest R-charges, $\langle \cO_{p_1} \bar{\cO}_{p_2} \bar{\cO}_{p_3} \cO_{p_4}\rangle$, where say $p_4$ is the largest weight. Then the degree of extremality is defined to be $\mathsf{k}=\tfrac12(p_1+p_2+p_3-p_4)$~\cite{DHoker:2000xhf} (see also~\cite{Aprile:2020uxk}). Recalling that we take $m>n$, the supergravity method of the present work would yield the above correlator with 
$p_1=n$, $p_2=m$, $p_3=p$
general, and $p_4=p+m-n$, 
which has degree of extremality $\mathsf{k}=n$.

The remainder of this paper is structured as follows. In Section~\ref{llm-setup} we describe the derivation of the closed-form LLM metric for the profile \eqref{eq:ripplon-prof-multi-intro} to quadratic order, and its conversion into de Donder-Lorentz gauge. In Section~\ref{sec:probe-calcs} we solve the minimally coupled massless scalar wave equation with appropriate boundary conditions, to obtain the two sequences of correlators of two CPOs and two descendants in position space. In Section~\ref{sec:mellin-1} we write the correlators in Mellin space, analyze the respective superconformal Ward identities, confirm the CFT expression of~\cite{Aprile:2017xsp,Rastelli:2016nze,Rastelli:2017udc} for the sequence $\langle \cO_3 \bar\cO_3 \bar\cO_{p} \cO_{p}\rangle$, present our compact explicit expression for the dynamical part of the sequence $\langle \cO_2 \bar\cO_3 \bar\cO_{p} \cO_{p+1}\rangle$, and confirm that it agrees with the results of~\cite{Rastelli:2016nze,Rastelli:2017udc,Arutyunov:2018neq,Arutyunov:2018tvn,Caron-Huot:2018kta}. In Section~\ref{sec:discussion} we discuss our results.

\section{Background fields}
\label{llm-setup}

\subsection{Asymptotically AdS$_5 \times $S$^5$ LLM solutions}

The asymptotically AdS$_5 \times $S$^5$ LLM solutions involve only the metric and five-form field strength. We shall focus primarily on the metric, whose general local form is given by~\cite{Lin:2004nb}:
\begin{align}
     ds^2 &\;=\; -h^{-2}(dt+V_i dx^i)^2 + h^2(dy^2+dx^i dx^i) + y e^G d\Omega_3^2 +y e^{-G} d\medtilde{\Omega}_3^2 \;, \nonumber \\
    h^{-2} &\;=\; 2y \cosh{G}\;, \qquad z=\frac{1}{2} \tanh{G}\;, \\ 
    y\partial_y V_i &\;=\; \epsilon_{ij} \partial_j z \;, \qquad y\left(\partial_i V_j - \partial_j V_i\right) = \epsilon_{ij} \partial_y z\;, \nonumber 
\end{align}
where $i=1,2$. The quantities $z$ and $V_i$ are given by
\begin{align}
\begin{aligned}
   z(x_1,x_2,y) &\,=\, \frac{y^2}{\pi} \int_{R^2} \frac{z(x'_1,x'_2,0)dx'_1 dx'_2}{((x-x')^2+y^2)^2}\;,\\
   V_i(x_1,x_2,y) &\,=\, \frac{\epsilon_{ij}}{\pi} \int_{R^2} \frac{z(x'_1,x'_2,0) (x_j-x'_j) dx'_1 dx'_2}{((x-x')^2+y^2)^2}\;.
\end{aligned}
\end{align}
The value of the function $z$ at $y=0$ is required to be $\pm 1/2$, in order for one or the other of the two three-spheres to shrink smoothly. This boundary condition is typically represented by a colouring of the $\mathbb{R}^2$ at $y=0$ into black ($z=+1/2$) and white ($z=-1/2$) regions. The total area of all the black regions is fixed. The black regions have an interpretation as `droplets' in a phase space of free fermions, which is holographically related to the free-fermion description of the dual 1/2-BPS operators of $\cN=4$ SYM~\cite{Corley:2001zk,Berenstein:2004kk}. 

In this work we shall take the profile function to be given by
\be
\label{eq:ripplon-prof-multi}
	r(\tilde\phi)\,=\, \sqrt{1+\alpha_1  \cos (n \tilde\phi )+\alpha_2 \cos (m \tilde\phi )}\;,
\ee 
where, as mentioned in the introduction, we have $|\alpha_1|+|\alpha_2|<1$, we take $m > n \ge 2$, and where $(r,\tilde\phi)$ are plane polar coordinates, related to $x_i$ via $x_1=r\cos\tilde{\phi}$ and $x_2=r\sin\tilde{\phi}$. When the $\alpha_i$ are small, this profile describes a two-ripple deformation of a unit circle. 
We shall eventually focus on $n=2$, $m=3$, however in much of what follows we shall keep $m,n$ general.

At linear order in $\alpha_1$ and $\alpha_2$, the supergravity fields are simply given by the sum of the fields corresponding to each of the linearized modes $m$ and $n$. At order $\alpha_1^2$ and $\alpha_2^2$, there are terms with frequencies $2m$ and $2n$, as well as zero-mode terms. The main novelty with respect to a single-mode profile (e.g.~with $\alpha_2=0$) is that at order $\alpha_1\alpha_2$, there is an interaction between modes that gives rise to terms in the supergravity fields with frequencies $m+n$ and $m-n$.

\subsection{Expansion of the background solution to quadratic order}

We work perturbatively in small $\alpha_1$, $\alpha_2$, to quadratic order. We expand the metric as
\be
\label{eq:g-expansion}
     g =g^{(0)} + g^{(1)} + g^{(2)} \,,
\ee
where
\be
\label{eq:g-expansion-2}
     g^{(1)} = \alpha_1\;\! g^{(1,0)} + \alpha_2 \;\! g^{(0,1)} \,, 
     \qquad g^{(2)} = \alpha_1^2 \;\! g^{(2,0)}+ \alpha_1\alpha_2\;\! g^{(1,1)}
     +\alpha_2^2\;\! g^{(0,2)}\,, 
\ee
and similarly for the five-form field strength. 

First, when $\alpha_1=\alpha_2=0$, the profile is a circle, and the background reduces to empty global AdS$_5 \times$S$^5$. The quantities $z$ and $V$ take the form~\cite{Lin:2004nb} 
\begin{align}
    z &\;=\; -\frac{r^2+y^2-1}{2 \sqrt{(r^2+1+y^2)^2-4r^2}}\;,\nonumber\\
    V_{\tilde\phi} &\;=\; \frac{1}{2} \left(1-\frac{r^2+y^2+1}{\sqrt{(r^2+1+y^2)^2-4r^2}}\right), \qquad V_r \;=\; 0\;.
\end{align}
After changing coordinates via
\be
y\;=\;R\cos\theta\,,\qquad r\;=\;\sqrt{R^2+1}\sin\theta\,,\qquad \tilde\phi \,=\, {\phi}-t\,,
\ee
one obtains the line element and five-form field strength of empty global AdS$_5\times$S$^5$ in the following form,
\be
\label{eq:metric0}
ds^2 \;=\;
-(R^2+1)dt^2 + \frac{dR^2}{R^2+1}+R^2 d\medtilde{\Omega}_3^2
+ d\theta^2 + \sin^2 \theta d\phi^2 + \cos^2\theta d\Omega_3^2 
\,,
\ee
\be
    F_5 \;=\; R^3\, dt\wedge dR \wedge d\medtilde{\Omega}_3 + \cos^3\theta\sin\theta\, d\theta \wedge d\phi \wedge d{\Omega}_3\,.
\ee

At linear order, the LLM profile~\eqref{eq:ripplon-prof-multi} gives a superposition of two linearised supergravitons in otherwise empty global AdS$_5\times$S$^5$. 
Upon deriving the fields from the profile \eqref{eq:ripplon-prof-multi}, one encounters negative powers of the quantity
\be
\label{eq:sigma}
\Sigma \,=\, R^2 + \cos^2\theta \;.
\ee
As done in~\cite{Grant:2005qc,Turton:2024afd}, we apply a diffeomorphism to convert the linear-order fields into the form of the Kaluza-Klein fluctuation analysis of~\cite{Kim:1985ez}, with a finite sum over S$^5$ harmonics. 
The vector field  $\xi^{(1)}$ that generates this diffeomorphism is the sum of two parts, corresponding to each of the linearized modes. The part of $\xi^{(1)}$ corresponding to a single mode is given in~\cite{Grant:2005qc,Turton:2024afd}.
The resulting linearised metric is in de Donder-Lorentz gauge, defined by (we denote AdS$_5$ indices by $\mu,\nu,\ldots$ and S$^5$ indices by $a,b,\ldots\,$, and we use round brackets on indices to denote the symmetric traceless part)
\be
    D^a h_{(ab)} \,=\, D^a h_{a\mu} \,=\, 0 \,.
\ee

The $(1,0)$ parts of the linearised metric and four-form potential are as follows (the $(0,1)$ parts are entirely analogous):
\be
    g^{(1,0)}_{\mu\nu} \,=\, \sum_{l=\pm n} \left(-\frac{6}{5} |l| s_l Y_l \, g^{(0)}_{\mu\nu} + \frac{4}{|l|+1} Y_l \nabla_{(\mu} \nabla_{\nu)} s_l \right), \qquad g^{(1,0)}_{\alpha\beta} \,=\, \sum_{l=\pm n} 2|l|s_l Y_l \, g^{(0)}_{\alpha\beta}\;,
    \label{eq:g-1-n}
\ee
\be
    A_4^{(1,0)} \,=\, \sum_{l=\pm n} \left( Y_l \star_{AdS_5} ds_l - s_l \star_{S^5} dY_l\right)\,,      \nonumber
\ee
where
\be
\label{eq:sn-Yn-def}
    s_l \;=\; \frac{|l|+1}{8|l|(R^2+1)^{|l|/2}} e^{i l t}\,, \qquad Y_l \;=\; e^{i l \phi} \sin^{|l|}\theta\,.
\ee
The functions $s_l$ and $Y_l$ are scalar harmonics on AdS$_5$ and S$^5$ respectively, satisfying
\be
\label{eq:s-Y}    \square_{\mathrm{\sst A}} s_l \,=\, l(l-4)s_l\,, \qquad \square_{\mathrm{\sst S}} Y_l \,=\, -l(l+4)Y_l\,, \qquad\quad l >0 .
\ee

At quadratic order, we first compute the closed-form fields that follow directly from the profile~\eqref{eq:ripplon-prof-multi}, specialising now to $n=2, m=3$.
The resulting (lengthy) expressions again contain negative powers of $\Sigma$. So we again construct a diffeomorphism to convert the second-order metric and five-form field strength into de Donder-Lorentz gauge. 
Specifically, after taking account of the second-order terms arising from the  $\xi^{(1)}$ diffeomorphism described below \eqref{eq:sigma}, we apply an additional quadratic order diffeomorphism generated by another vector field $\xi^{(2)}$, which brings both metric and five-form into the desired form. 
Due to their length, we shall not write out the explicit quadratic order fields in the current work; instead, we record the explicit form of $\xi^{(2)}$ in Appendix \ref{app:metric-quad-diffeo}. 
We emphasize that although we focus on $n=2, m=3$ in the present work, this procedure can be carried out systematically for any specific choice of values of $m$ and $n$.

\section{Probing the solutions with a massless scalar}
\label{sec:probe-calcs}

\subsection{Equations of motion and boundary conditions}

The linearized fluctuation of the dilaton/axion satisfies the 10D minimally coupled massless scalar wave equation on the curved background we study,
\be
\label{eq:box-phi-full-10}
    \square \Phi \,=\, 0\,.
\ee
We expand all quantities perturbatively, using the notation 
\be
\label{eq:pert-exp-012}
     \Phi \;=\;\Phi^{(0)} + \Phi^{(1)} + \Phi^{(2)} \,, \qquad\qquad 
     \square \;=\; \square^{(0)} + \square^{(1)} +\square^{(2)} \,,
\ee
and similar, where the $\alphaone$ and $\alphatwo$ dependence of the first- and second-order terms is analogous to that of the metric, as given in \eqref{eq:g-expansion-2}.
Expanding the equation of motion \eqref{eq:box-phi-full-10}, one obtains the set of perturbative equations
\begin{align}
    \square^{(0)} \Phi^{(0)} &\;=\; 0\,,
    \label{eq:eom-0} \\
    \square^{(0)} \Phi^{(1)} &\;=\; -\square^{(1)}\Phi^{(0)}\,,\label{eq:eom--1}\\
    \square^{(0)} \Phi^{(2)} &\;=\; -\square^{(2)}\Phi^{(0)}-\square^{(1)}\Phi^{(1)}\,,
    \label{eq:eom-2}
\end{align}
which are further divided into equations controlled by separate expansions in $\alphaone$ and $\alphatwo$.

The scalar field $\Phi$ expands in scalar spherical harmonics on S$^{5}$, 
\be
\label{eq:B-I-expand}
    \Phi(x,y) \,=\, \sum\limits_{I} B_{I}(x) Y^{I}(y) \,,  
\ee
where from now on $x$ denotes a point in AdS$_5$ and $y$ a point in S$^5$. 
We expand the coefficients $B_{I}$ in the way indicated in Eq.~\eqref{eq:pert-exp-012}.

Let us now describe the boundary conditions that are appropriate for the families of correlators of interest to us. 
For the first sequence of correlators, $\langle \mathcal{O}_3 \bar{\mathcal{O}}_3 \bar{\mathcal{D}}_{k}  \mathcal{D}_{k}  \rangle$, it suffices to consider the background in which the amplitude of one of the modes is set to zero, say $\alphaone=0$, and $m=3$. 
Our calculation does not distinguish between a dilaton or an axion fluctuation, since both satisfy~\eqref{eq:box-phi-full-10}. When interpreting the correlator in the dual CFT, we will focus on operators that are U(1) hypercharge eigenstates, i.e.~$\mathcal{D}_k  \sim Q^4 \cO_{k+2}$ and $\bar{\mathcal{D}}_k  \sim \bar{Q}^4 \bar\cO_{k+2}$. These operators have dimension $\Delta = k+4$ and have $R$-charge $J=\pm k$. The corresponding boundary condition involves a source in a highest/lowest-weight harmonic of degree $k$, $Y_{\pm k}$. We seek a solution with a source at position $\vec{n}$ on the boundary, with a response in the same harmonic, and that is everywhere regular,
\be
\label{eq:bc-33kk}
    \lim_{R\to \infty} \Phi  \;=\; \frac{\delta(\vec{x}-\vec{n}) Y^{I}(y)}{R^{4-\Delta}} + \frac{b_{I}(\vec{x}) Y^{I}(y)}{R^{\Delta}} +\cdots\,,
\ee
where $Y^I=Y_{-k}$ for concreteness. Here $\vec{x}$ denotes a boundary coordinate in $\mathbb{R}\times S^3$, $b_I(\vec{x})$ stands for the response proportional to the harmonic $Y^I$, and the dots denote response terms proportional to other harmonics. 
Then the response, which arises at order $\alphatwo^2$, encodes the dynamical part of the following LLLL correlator in the supergravity regime,
\be
b_{k}^{[1]}(\vec{x}) \;=\; \alphatwo^2
    \big\langle \mathcal{O}_3 (0) \bar{\mathcal{O}}_3 (\infty) \bar{\mathcal{D}}_{k} (\vec{n}) \mathcal{D}_{k} (\vec{x}) 
    \big\rangle 
    \;,
\ee
where as usual $ \langle {\mathcal{O}}_\Delta (\infty) \cdots \rangle \equiv \lim_{|\vec{x}|\to \infty}|\vec{x}|^{2\Delta}\langle {\mathcal{O}}_\Delta (\vec{x}) \cdots \rangle\,$. For later convenience we have introduced the superscript $[1]$ to denote Sequence 1.

For the second sequence, 
$\langle \mathcal{O}_2 \bar{\mathcal{O}}_3  \bar{\mathcal{D}}_{k} \mathcal{D}_{k+1} \rangle$, we specify the source in the same harmonic as before, $Y^I=Y_{-k}$, but instead focus on the response in the lowest-weight harmonic with one degree higher than the source, i.e.
\be
\label{eq:bc-kkp1}
    \lim_{R\to \infty} \Phi  \,=\, \frac{\delta(\vec{x}-\vec{n}) Y^{I}(y)}{R^{4-\Delta}} + \frac{b_{J}(\vec{x}) Y^{J}(y)}{R^{\Delta}} +\cdots\,,
\ee
where $Y^{J} = Y_{-(k+1)}$. This response arises at order $\alphaone\alphatwo$, and it encodes the dynamical part of the following LLLL correlator in the supergravity regime, 
\be
\label{eq:seq-2-corr}
b_{k+1}^{[2]}(\vec{x}) \;=\; \alphaone\alphatwo
    \big\langle \mathcal{O}_2 (0) \bar{\mathcal{O}}_3 (\infty) \bar{\mathcal{D}}_{k} (\vec{n}) \mathcal{D}_{k+1} (\vec{x}) 
    \big\rangle 
    \;.
\ee

\subsection{Solving the equations up to linear order}

Since there are several common elements, we shall describe the calculations for the two sequences of correlators together. The common first step is that at zeroth order, we specify as input the solution to \eqref{eq:eom-0} with respective boundary conditions \eqref{eq:bc-33kk} and \eqref{eq:bc-kkp1}. In both cases, the solution is
\be
\label{eq:phi--0}
  \Phi^{(0)}(x,y) \,=\, \hat{B}_{k+4}(x) Y_{-k}(y) \,,
\ee
where $\hat{B}_{\Delta}$ is the bulk-to-boundary propagator of dimension $\Delta$ with boundary point $\vec{n}$.

In global AdS$_5$, we denote coordinates by $\vec{R}=(t,R,\vec{\Omega})$. On the conformal boundary $\mathbb{R}\times$S$^3$, we use adapted coordinates such that the specified boundary point $\vec{n}$ is at $t=0$ and the North pole of S$^3$ ($\beta=0$ in what follows). 
On the S$^3$ in AdS$_5$, 
we use hyperspherical coordinates, 
such that the line element is
\begin{align}
d\medtilde\Omega_3^2 \,&=\, d\beta^2 +\sin\beta^2 ( d\gamma^2 + \sin\gamma^2 d\delta^2 ) \;.
\end{align}
In these coordinates, $\hat{B}_{\Delta}$ takes the form
\be
\label{eq:bhat--1}
\hat{B}_{\Delta}(\vec{R}) \;\equiv\; \left( \frac12
\frac{1}{\sqrt{R^2+1}\cos t - R \cos \beta} \right)^\Delta \;.
\ee

At linear order, we seek the general solution to the equation of motion~\eqref{eq:eom--1} for $\Phi^{(1)}$ that will eventually contribute to the dynamical part of the correlators of interest. By linearity, we have
\be
		\Phi^{(1)} \,=\, \alpha_1 \;\! \varphi^{(1,0)} + \alpha_2 \;\! \varphi^{(0,1)} \,,
        \qquad
        \square^{(1)} \,=\, \alpha_1 \;\! \square^{(1,0)} + \alpha_2 \;\! \square^{(0,1)} \,,
\ee    
with 
\be
\label{eq:1-0-eom}
\square^{(0)} \varphi^{(1,0)} = -\square^{(1,0)} \Phi^{(0)}\,,
\ee
and similarly for $\varphi^{(0,1)}$. We next solve \eqref{eq:1-0-eom} for general $n$, then later superpose as needed. For ease of presentation, we temporarily set $\alpha_2=0$,
and work with $\Phi^{(1)}$ rather than $\varphi^{(1,0)}$. Moreover, we also suppress factors of $\alpha_1$ throughout most of what follows, and restore them where appropriate.

To solve \eqref{eq:1-0-eom} for general $n$, we note that the right-hand side involves the linear order background~\eqref{eq:g-1-n}). 
One can simplify the right-hand side considerably by making an auxiliary diffeomorphism which sets to zero $g^{(1)}_{tR}$ and introduces instead a non-zero $g^{(1)}_{\theta\phi}$, as done in~\cite{Giusto:2024trt} for $n=2$. For general $n$, the linearised diffeomorphism is generated by the following vector field:
\be
\label{eq:diff-gen}
   (\xi')^{M} 
    \,=\, 
    -\frac{2}{n+1}\Big( Y_n \nabla^{M} s_n 
       - s_n \nabla^{M} Y_n 
       \,+\,c.c.
    \Big)\,,
\ee
where $M$ is a ten-dimensional index.

To write the simpler form of Eq.\;\eqref{eq:eom--1}, we note that due to the symmetries of the background, in the Kaluza-Klein expansion of the fields in S$^5$ 
harmonics, only SO(4)-invariant $S^5$ spherical harmonics appear in the analysis, whereupon the multi-index $I$ reduces to $I=(\mathsf{l},\mathsf{m})$; see for instance~\cite{Skenderis:2006uy,Turton:2025svk}. Then, after the diffeomorphism generated by \eqref{eq:diff-gen}, the linearised equation~\eqref{eq:eom--1} becomes
\be
    \square^{(0)} \Phi^{(1)'} \,=\, 
    \begin{cases}
   \qquad\qquad 0 &~~~ k=0 \;, \\[2mm]
 \displaystyle  \qquad -\frac{32 n (n-1) }{n+1} s_n \hat{B}_{5} Y_{n-1} & ~~~k =1 \;, \label{eq:o-a-eom-n-g} \\[4mm]
\displaystyle \frac{16 n (n-1) k (k-1)}{n+1} s_n \hat{B}_{k+4} Y^{(k,2-k)} Y_{n-2} & ~~~k \ge 2 \;. \vspace{3mm}
    \end{cases}
\ee
We start with the generic case, $k\ge n \ge 2$. Since the source term contins a product of a highest-weight spherical harmonic and an SO(4)-invariant descendant harmonic, we make an Ansatz that $\Phi^{(1)'}$ consists of a sum of similar products of harmonics, 
with total U(1) R-charge $n-k$. This Ansatz is based on identities for $\square^{(0)}$ acting on such products, as described in Appendix~\ref{app:recur-sol}. Explicitly, the Ansatz is
\be
   \qquad  \Phi^{(1)'}(x,y) \,=\, \sum_{i=0}^{n-2} \Phi_{(i)}(x) Y^{(k-i,2+i-k)}(y) Y_{n-i-2}(y) \;.
\ee
This gives rise to a recursive system of equations for the $\Phi_{(i)}\:\!$; the derivation of this system, and details of our approach to solving it, are given in Appendix \ref{app:recur-sol}. The resulting solution for the $\Phi_{(i)}$ for $i=0,\ldots, n-2$ is
\be
\label{eq:phi-i-sol}
   \Phi_{(i)}(x) \,=\, -\frac{4(n-i-1)}{n-i}\frac{(k-i+2)(k-i+1)(k-i)(k-i-1)}{(k+3)(k+2)(k+1)} s_{n-i-1}(x) \hat{B}_{k-i+3}(x) \;.
\ee

Here we have discarded the part of the general solution arising from the corresponding homogeneous equations, since this does not give any contribution to the dynamical part of the correlators we study. 
The above formulae give the general solution $\Phi^{(1)'}$ for any $k \ge n\geq2$ that will eventually contribute to the dynamical part of these correlators.

Let us now discuss the case $k < n$. For $k=0$ the equation~\eqref{eq:o-a-eom-n-g} is homogeneous, and we take the trivial solution. For $1 \le k <n$, the solutions can be efficiently derived from careful extrapolations of the solution for $k\ge n$, as described in Appendix~\ref{app:recur-sol}. 
In particular, for $k=1$ and general $n$, we obtain the compact expression
\be
\label{eq:k1sol}
\Phi^{(1)'}_{k=1} \,=\, 
\frac{2(n-1)}{n}s_{n-1}(x)
\hat{B}_{4}(x)  Y_{n-1}(y) \,.
\ee
More generally, for $1 \le k <n$, the solution is a sum of $k$ terms.
The expressions \eqref{eq:phi-i-sol} and \eqref{eq:k1sol} agree with the solution for $n=2$ obtained in~\cite{Turton:2024afd}.

We next transform back to de Donder-Lorentz gauge. Upon doing so, the scalar probe picks up additional terms of the form
\be
\label{eq:sol-ddl-ga-pre}
    \delta\Phi^{(1)} \,=\, \frac{2}{n+1}\Big( (Y_n \nabla^{M}s_n-s_n \nabla^{M}Y_n) + c.c. \Big)\partial_M \Phi^{(0)}  \,,
\ee
such that
\be
\label{eq:sol-ddl-ga}
   \Phi^{(1)} \,=\, \Phi^{(1)'} +\delta\Phi^{(1)}\,.  
\ee
All together, the linear-order solution $\Phi^{(1)}$ on the background defined by the two-mode profile \eqref{eq:ripplon-prof-multi} for general $m,n$ is given by the sum of the solutions for the respective modes,
\be
\label{eq:lin-sol-gen}
		\Phi^{(1)} \,=\, \alpha_1 \;\! \varphi^{(1,0)}_{n} + \alpha_2 \;\! \varphi^{(0,1)}_{m}\;.
\ee

\subsection{Linear order solution and quadratic source for Sequence 1}

For the first sequence of correlators,  $\langle \mathcal{O}_3 \bar{\mathcal{O}}_3 \bar{\mathcal{D}}_{k}  \mathcal{D}_{k} \rangle$, we specialize the background and linear order solution \eqref{eq:lin-sol-gen} to $m=3,\; \alpha_1=0$. Suppressing an overall factor of $\alpha_2$, the $k\ge 2$ solution in the transformed gauge is (for details, see Appendix~\ref{app:recur-sol}):
\begin{align}
\begin{aligned}
\Phi^{(1)'}_{k=2} \,&=\, -\frac{16}{15} s_2(x) \hat{B}_5(x) Y^{(2,0)}(y) Y_1(y) + \frac{2}{5} s_1(x) \hat{B}_4(x) Y_1(y) \,, \label{eq:phi1p-tt3-new} \\[2mm]
\Phi^{(1)'}_{k\ge 3} \,&=\, -\frac{8}{3}\frac{k(k-1)}{k+3} s_{2}\hat{B}_{k+3} Y^{(k,2-k)} Y_1-\frac{2k(k-1)(k-2)}{(k+3)(k+2)} s_1 \hat{B}_{k+2} Y^{(k-1,3-k)} \,,
\end{aligned}
\end{align}
and the full solution for $\Phi^{(1)}$ is obtained by combining this with Eqs.\;\eqref{eq:k1sol}, \eqref{eq:sol-ddl-ga-pre} and \eqref{eq:sol-ddl-ga} for $n=3$.

We then compute the source of the quadratic order equation of motion \eqref{eq:eom-2}.  It is convenient to write this
equation in the form
\be
\label{eq:box-phi-2-J}
    \qquad \qquad \qquad
    \square^{(0)}\Phi^{(2)} \,=\, \mathcal{J}\,, \qquad \quad \cJ\,\equiv\,-\square^{(2)}\Phi^{(0)}-\square^{(1)}\Phi^{(1)} \,.
\ee
To isolate the solution $B_I^{(2)}$ of interest, we project the source $\cJ$ on the appropriate spherical harmonic. For the first sequence of correlators, $\langle \mathcal{O}_3 \bar{\mathcal{O}}_3 \bar{\mathcal{D}}_{k}  \mathcal{D}_{k} \rangle$, we must project on $Y_{-k}$,
\be
\label{eq:Jk-def}
\mathcal{J}^{[1]}
\,\equiv \,
 \frac{1}{||Y_{k}||^2}  \int d\Omega_5 \,Y_{-k}^{*}\,
    \mathcal{J}\big|_{\alpha_1=0}
     \;,
\ee
where we recall that $[1]$ denotes Sequence 1. 
We thereby obtain the following projected five-dimensional equation of motion to solve,
\begin{align}
\qquad\qquad\qquad
    \square_{\mathrm{\sst A}}^{(0)} B_k^{(2)} -m_1^2 B_k^{(2)}
    &{}\;=\;
    \mathcal{J}^{[1]} 
    \;, \qquad \quad m_1^2 \,=\, k(k+4)  \,.
    \label{eq:proj-sources-gen}
\end{align}

Before proceeding to solve this equation, we next describe the analogous equation for the Sequence 2 correlators.


\subsection{Linear order solution and quadratic source for Sequence 2}

For the second sequence, $\langle \mathcal{O}_2 \bar{\mathcal{O}}_3  \bar{\mathcal{D}}_{k} \mathcal{D}_{k+1} \rangle$, we use the full background with both the modes $n=2$ and $m=3$. So the linear-order solution for the probe is given by
\be
		\Phi^{(1)} \,=\, \alpha_1 \;\! \varphi^{(1,0)}_{n=2} + \alpha_2 \;\! \varphi^{(0,1)}_{m=3}\;.
\ee
The solution for $n=2$ and $k \ge 2$ is~\cite{Turton:2024afd}
\be
\label{eq:phi1pkge2-tt1}
    \Phi^{(1)'}_{k\ge 2} \,=\, - \frac{2k(k-1)}{k+3} \hat{B}_{k+3}(x) s_{1}(x) Y^{(k,2-k)}(y)\,.
\ee
Following the same method, we compute the source of the quadratic equation of motion. However, this time we must project on $Y_{-(k+1)}$.  We thus write the projected source for the  correlators in Sequence 2 as 
\be
\label{eq:Jkp1-def}
\mathcal{J}^{[2]}
\,\equiv \,
 \frac{1}{||Y_{k+1}||^2}  \int d\Omega_5 \,Y_{-(k+1)}^{*}\,
    \mathcal{J}
  \; .
\ee
The projected five-dimensional equation for Sequence 2 is then
\begin{align}
\qquad \qquad \qquad 
    \square_{\mathrm{\sst A}}^{(0)} B_{k+1}^{(2)} -m_2^2 B_{k+1}^{(2)}
    &{}\;=\;
    \mathcal{J}^{[2]}
    \;, \qquad \quad  m_2^2 \,=\, (k+1)(k+5) \,.
    \label{eq:proj-sources-gen-2}
\end{align}

\subsection{Solving the quadratic order equation of motion}

For both families of correlators, we next solve the projected five-dimensional quadratic-order equations of motion~\eqref{eq:proj-sources-gen} and~\eqref{eq:proj-sources-gen-2}  using the Green's function, i.e.~the bulk-to-bulk propagator $G^{\mathrm{Glob}}_{\Delta}(\vec{R}'|\vec{R})$, 
\be
\label{eq:corr-sug-5d}
    B^{(2)}_I (\vec{R}) \;=\; 
    i \! \int d^5 \!\!\; \vec{R}' \sqrt{-g_{AdS_5}} \, G^{\mathrm{Glob}}_{\Delta}(\vec{R}'|\vec{R}) \, 
    \mathcal{J}^{[i]}(\vec{R}') \;,
\ee
where here $i=1,2$, and where for Sequence 1, $\Delta=k+4$, and for Sequence 2, $\Delta=k+5$.

Then, to derive four-point correlators, we examine this solution near asymptotic infinity in AdS$_5$, whereupon the bulk-to-bulk propagator becomes a bulk-to-boundary propagator.
More explicitly, at leading order in large $R$, the bulk-to-bulk propagator 
$G^{\mathrm{Glob}}_{\Delta}$ is related to the bulk-to-boundary propagator $K^{\mathrm{Glob}}_{\Delta}$ of global AdS$_5$ as follows: 
\be
\lim_{R \to \infty}
    G^{\mathrm{Glob}}_{\Delta}(\vec{R'}|R,t,\vec{\Omega}) ~\,=\,~ \frac{\Gamma(\Delta)}{2 \pi^{2} \Gamma(\Delta-1)} \:\! \frac{1}{R^{\Delta}} \:\! K^{\mathrm{Glob}}_{\Delta}(\vec{R'}|t,\vec{\Omega})\,,
\ee
where the bulk-to boundary propagator $K^{\mathrm{Glob}}_{\Delta}$ takes the form 
\be
\label{eq:k-glob-main}
    K^{\mathrm{Glob}}_{\Delta}(\vec{R'}|t,\vec{\Omega}) \;\equiv\; \left(\frac{1}{2}\frac{1}{\sqrt{1+R'^2} \, \cos(t-t')-R'\, \vec{\Omega}\cdot \vec{\Omega}'  }\right)^{\Delta}\,.
\ee
Here, the notation $\vec{\Omega}$ labels a point on the S$^3$ inside AdS$_5$ via a unit-normalized vector in $\mathbb{R}^4$, such that\footnote{We suppress the tilde on $\vec{\Omega}$, since we shall not need similar notation for the three-sphere in S$^5$.} 
\be
    \vec{\Omega}\cdot \vec{\Omega}' \;=\; \sin\beta \sin\beta' \left(\sin\gamma\sin\gamma'\cos(\delta-\delta')+\cos\gamma\cos\gamma'\right) + \cos\beta \cos\beta'\,.    
\ee
The response is then given by
\be
\label{eq:resp-hyp-10}
    b_I(t,\vec{\Omega}) \,=\,  {}  \frac{\Gamma(\Delta)}{2 \pi^{2} \Gamma(\Delta-1)} \!\; i \!\!\!\; \int d^5 \vec{R}' \sqrt{-g_{AdS_5}} \, K^{\mathrm{Glob}}_{\Delta}(\vec{R'}|t,\vec{\Omega}) \, \mathcal{J}^{[i]}(\vec{R}')
     \,.
\ee

Our next goal is to obtain expressions for the two sequences of correlators in terms of the usual $D$-functions~\cite{DHoker:1999kzh}. Therefore, we now rewrite the quadratic sources as sums of products of bulk-to-boundary propagators.
At this point, we Wick rotate
to Euclidean time $t_e=i t$, we denote Euclidean
hyperspherical coordinates as $\vec{R}_e=(t_e,R,\vec{\Omega})$, and we define
\begin{align}
\begin{aligned}
\label{eq:b-to-bdy-hy}
B^+_\Delta(\vec{R}_e) \;&\equiv\, 
\left(\frac{e^{t_e}}{\sqrt{1+R^2}}\right)^{\Delta} \,,
\qquad\qquad
B^-_\Delta(\vec{R}_e) \;\equiv\,
\left(\frac{e^{-t_e}}{\sqrt{1+R^2}}\right)^{\Delta}\,,
\\
\hat{B}_\Delta(\vec{R}_e) \;&\equiv\, 
\left( \frac12
\frac{1}{\sqrt{R^2+1}\cosh t_e - R \cos \beta} \right)^\Delta \;,
\end{aligned}
\end{align}
where $\hat{B}_\Delta$ is obtained from the Wick rotation of Eq.~\eqref{eq:bhat--1}.

When the sources $\cJ^{[1]}$, $\cJ^{[2]}$ are written as a sum of products of these functions, the terms that contribute to the dynamical part of four-point correlators involve products of precisely one of each of $B^+_{\Delta_1}\,$, $B^-_{\Delta_2}$ and $\hat{B}_{\Delta_3}$ for some $\{\Delta_i\}$. We therefore discard all other terms.
The coefficients in this sum depend only on $k$. 
We record the explicit form of this dynamical part of $\cJ^{[1]}$ and $\cJ^{[2]}$ in Appendix~\ref{app:sources-both}.


\subsection{Two sequences of four-point correlators in terms of D-functions}


In order to express the correlators in terms of $D$-functions, we change to Euclidean Poincar\'e coordinates $\boldsymbol{w} \equiv (w_0,\vec{w})$ as follows,
\begin{align}
    w_0 \;=\; \frac{e^{t_e} }{\sqrt{1+R^2}}\;, \qquad
    w_1 &\;=\; w_0 R \sin \beta \sin \gamma \cos \delta \;, \qquad
    w_2 \;=\; w_0 R \sin \beta \sin \gamma \sin \delta \;,\nonumber \\
    w_3 &\;=\; w_0 R \sin \beta \cos \gamma  \;,\qquad
    w_4 \;=\; w_0 R \cos \beta \;,
\end{align}
whereupon the line element becomes
\be
    ds_{\mathrm{\sst{EAdS_5}}}^2 \,=\,  \frac{1}{w_0^2}\bigg( dw_0^2 + \sum\limits_{i=1}^{4} dw_i^2 \bigg)\;.
\ee

In the remainder of this work, we denote the coordinates on the boundary $\mathbb{R}^4$ at $w_0=0$ by $\vec{x}=(x^1,x^2,x^3,x^4)$. 
These coordinates are related to those on the Euclidean cylinder $\mathbb{R}\times $S$^3$ at $R = \infty$ by
\be 
    x^1 = e^{t_e} \sin \beta \sin \gamma \cos \delta \;, ~~
    x^2 = e^{t_e} \sin \beta \sin \gamma \sin \delta \;, ~~
    x^3 = e^{t_e} \sin \beta \cos \gamma \;, ~~
    x^4 = e^{t_e} \cos \beta \;.
\ee

The bulk-to-boundary propagators with boundary points at $(w_0=0,\vec{x})$ and at $w_0=\infty$ take the following respective forms in Euclidean Poincar\'e coordinates~\cite{Witten:1998qj},
\begin{align}
\label{eq:k-ads5-poin-1}
K_\Delta(\boldsymbol{w}\:\!|\:\!\vec{x}) \;&\equiv\; \left(\frac{w_0}{w_0^2+|\vec{w}-\vec{x}|^2}\right)^{\Delta}\;,
\qquad 
K_\Delta(w_0\:\!|\:\!\infty) \;\equiv\; w_0^\Delta \;.
\end{align}
Furthermore, the relation between the bulk-to-boundary propagators in the two coordinate systems is
\be
    \label{eq:k-glob-k-poin-1}
    K^{\mathrm{Glob}}_{\Delta}(\vec{R}'_e|t_e,\vec{\Omega}) \;=\; |\vec{x}|^{\Delta} K_{\Delta}(\boldsymbol{w}|\vec{x})\;. 
\ee

Recall that we have written the sources in terms of three bulk-to-boundary propagators, which respectively have boundary points at $t_e=\pm \infty$, and the point $\vec{n}\in\mathbb{R}\times$S$^3$ defined by $\,t_e=\beta=0$.
Upon changing to Euclidean Poincar\'e coordinates,
the point at $t_e = \infty$ maps to $w_0 =\infty$; the point $t_e \to -\infty$ maps to the origin $\vec{x}=0$ at $w_0=0$; and the point $\vec{n}\in\mathbb{R}\times$S$^3$ 
corresponds to the point $\vec{x}=\vec{n}_x\equiv (0,0,0,1)$ at $w_0=0$.
Explicitly, we have the relations
\begin{align}
B^+_\Delta(\vec{R}_e) \,&=\, K_{\Delta}(\boldsymbol{w}|\infty) 
\,=\, \lim_{|\vec{x}|\to \infty}|\vec{x}|^{2\Delta}
K_{\Delta}(\boldsymbol{w}|\vec{x})
\,,
\nonumber\\
B^-_\Delta(\vec{R}_e) \,&=\, K_{\Delta}(\boldsymbol{w}|0) \,, \qquad \quad
\hat{B}_\Delta(\vec{R}_e) \,=\, 
    K^{\mathrm{Glob}}_{\Delta}(\vec{R}'_e|\vec{n})
    \,=\,
    K_{\Delta}(\boldsymbol{w}|\vec{n}_x)\,.
\end{align}

Upon changing to Euclidean Poincar\'e coordinates, the boundary maps from the cylinder $\mathbb{R}\times$S$^3$ to flat $\mathbb{R}^4$ (plus the point at infinity). 
The dynamical part of the four-point correlator on flat $\mathbb{R}^4$ is then given by 
\be
\label{eq:sugra-corr}
    b_I^{\mathrm{flat}}(\vec{x}) \;=\;  {} \frac{\Gamma(\Delta)}{2 \pi^{2} \Gamma(\Delta-1)} \int d^5 \boldsymbol{w'} \sqrt{\bar{g}} \, K_{\Delta}(\boldsymbol{w'}\:\! |\:\! \vec{x} ) \, \mathcal{J}^{[i]}(\boldsymbol{w'}) \;.
\ee

We are now in a position to express the holographic correlator in terms of $D$-functions, which we do using the conventions of~\cite{Turton:2024afd}, summarized in Appendix~\ref{app:D-fns-conventions}.
To do so, let us introduce some notation.
We consider CFT four-point correlators of scalars $O_i$ of respective scaling dimensions $\Delta_i$. 
We focus on correlators with $\Delta_2+\Delta_3\geq \Delta_1+\Delta_4$, which both Sequence 1 and Sequence 2 satisfy. Then conformal invariance implies that the correlator takes the form \cite{Rastelli:2017udc}
\be
\langle
O_1(x_1) O_2(x_2) O_3(x_3) O_4(x_4)
\rangle 
\,=\,
\cK_{\Delta_1\Delta_2\Delta_3\Delta_4}\,
\cG_{\Delta_1\Delta_2\Delta_3\Delta_4}
(U,V) \,,
\ee
where
\be
\label{eq:cK-defn}
    \cK_{\Delta_1\Delta_2\Delta_3\Delta_4} \,=\, 
      \frac{\left(x_{13}^2\right)^{\frac{1}{2} (-\Delta_1+\Delta_2-\Delta_3+\Delta_4)} \left(x_{23}^2\right)^{\frac{1}{2} (\Delta_1-\Delta_2-\Delta_3+\Delta_4)}}{\left(x_{12}^2\right)^{\frac{1}{2} (\Delta_1+\Delta_2-\Delta_3+\Delta_4)}\left(x_{34}^2\right)^{\Delta_4} }\;,
\ee
where $x_{ij}^2\equiv(\vec{x}_i-\vec{x}_j)^2$, and where $U,V$ are the conformal cross-ratios (for our conventions, see~\cite{Turton:2024afd})
\be
\label{eq:U-V-def}
    U \,\equiv\, \frac{x_{12}^2 x_{34}^2}{x_{13}^2 x_{24}^2}\,,\qquad
    V \,\equiv\, \frac{x_{14}^2 x_{23}^2}{x_{13}^2 x_{24}^2}\;.
\ee 
For later use, we denote the correlator with points $x_1=0$, $x_2\to\infty$, $x_3=\vec{n}_x$ and $x_4=\vec{x}$ being generic as $\mathcal{C}_{\{\Delta_i\}}$,
\be               
    \mathcal{C}_{\{\Delta_i\}}(U,V) \,\equiv\, \langle O_1 (0) \bar{O}_2 (\infty) \bar{O}_3 (\vec{n}_x) O_4 (\vec{x}) \rangle   \,. 
\ee

As is often done, we separate the supergravity correlator into a sum of its value in free $\cN=4$ SYM and its remaining dynamical part (see e.g.~\cite{Aprile:2017xsp}),
\be
\label{eq:g-dyna}
\cG^{(\mathrm{{sugra}})}_{\{\Delta_i\}}(U,V) \,=\, 
\cG^{(\mathrm{{free}})}_{\{\Delta_i\}}(U,V)
+\cG^{(\mathrm{{dyna}})}_{\{\Delta_i\}}(U,V) \,.
\ee
We focus in the remainder of this work on the dynamical part of the correlator in the supergravity regime, which is what the supergravity calculation yields. With this understood, we shall drop the label ${}^{(\mathrm{{dyna}})}$.

The outcome of the supergravity calculation is a lengthy expression for each of the two sequences of correlators. These expressions are not unique in position space, due to the existence of multiple identities between the various $D$-functions; for completeness, they are given in Appendix~\ref{app:sugra-corrs}. 
As we shall see, it is significantly more convenient to transform to Mellin space to analyse the correlators.


\section{Mellin space correlators and superconformal Ward identity}
\label{sec:mellin-1}

\subsection{Supergravity four-point correlators}

Let us introduce some notation for writing correlators in Mellin space. 
The Mellin space amplitude $\cM(s,t)$ corresponding to $\cG$ is defined as follows. Using the conventions of~\cite{Giusto:2019pxc,Turton:2024afd},
\begin{align}
\mathcal{G}_{\{\Delta_i\}}(U,V)\,=\, \frac{\pi^2}{2} &\int \frac{ds}{4\pi i} \frac{dt}{4\pi i} U^{\frac{s}{2}-\frac{\Delta_3-\Delta_4}{2}}
 V^{\frac{t}{2}-\frac{\Delta_1+\Delta_4}{2}} 
    \Gamma\left[\frac{\Delta_1+\Delta_2}{2}-\frac{s}{2}\right] 
    \Gamma\left[\frac{\Delta_3+\Delta_4}{2}-\frac{s}{2}\right] 
     \nonumber\\
    &{} \times \Gamma^2\left[\frac{\Delta_2+\Delta_3-t}{2}\right]       \Gamma\left[\frac{\Delta_1+\Delta_3-u}{2}\right]        \Gamma\left[\frac{\Delta_2+\Delta_4-u}{2}\right] 
    \mathcal{M}_{\{\Delta_i\}}(s,t)\,,   
\end{align}
where $u=\hat\Delta-s-t=2(\Delta_1+\Delta_4)-s-t\,$, where $\hat{\Delta}\equiv\sum_i \Delta_i\,$, and we used the fact that $\Delta_1+\Delta_4=\Delta_2+\Delta_3$ for 
both Sequence 1 and 2.

We now take the Mellin transform of these two sequences of supergravity four-point correlators, expressed in terms of $\hat{D}_{\Delta_1 \Delta_2 \Delta_3 \Delta_4}(U,V)$, see Eq.~\eqref{eq:d-hat-defn}.
The Mellin transform of the $\hat{D}$-functions is given by
\begin{align}
    \hat{D}_{\Delta_1 \Delta_2 \Delta_3 \Delta_4}(U,V) \,=\,\:
    &
    \frac{\pi^{2}}{2\prod\limits_{j=1}^4 
    \Gamma(\Delta_j)} \Gamma\bigg(\frac{\hat{\Delta}-4}{2}\bigg)
    \int \frac{ds}{4\pi i} \frac{dt}{4\pi i} U^{\frac{s}{2}} V^{\frac{t}{2}} \Gamma\left[-\frac{s}{2}\right] \Gamma\left[-\frac{t}{2}\right] 
    \cr
    &~~\;
    \times
    \Gamma\left[\Delta_4+\frac{s+t}{2}\right] 
    \Gamma\left[ \frac{\Delta_1+\Delta_2 -\Delta_3 - \Delta_4 -s}{2} \right] 
    \\
    &~~\;
    \times
    \Gamma\left[ \frac{\Delta_2+\Delta_3 -\Delta_1 - \Delta_4 -t}{2} \right] \Gamma\left[ \frac{\Delta_1+\Delta_3 +\Delta_4 - \Delta_2 + s +t }{2} \right]\;.~~\;~~
    \nonumber
\end{align}

Then, up to an overall numerical factor that does not depend on $k$, we obtain the following result for the Sequence 1 correlators $\langle \mathcal{O}_3 \bar{\mathcal{O}}_3 \bar{\mathcal{D}}_{k} \mathcal{D}_{k}  \rangle$ in Mellin space: %
\begin{align}
   &{}\mathcal{M}_{33,k+4,k+4} \,=\,
\scalemath{0.92}{
   \frac{1}{d_k^{[1]}}   \Bigg( \frac{\left(k^2+7 k+12\right) u^2-2 \left(k^3+11 k^2+48 k+72\right) u +k^4+15 k^3+99 k^2+329 k+436}{s-2} 
   }
 \nonumber\\
    &
    \scalemath{0.92}{
    {}+ \frac{\left(k^3+9 k^2+22 k+8\right) u^2-2 \left(k^4+11 k^3+56 k^2+120 k+40\right) u +k^5+13 k^4+93 k^3+355 k^2+666 k+200}{s-4}
    }
    \nonumber\\ \label{eq:33kk-sugra}
     &
     \scalemath{0.92}{
     {}\qquad\qquad\qquad +\frac{8 (k-1) k^2}{u-(k+3)} +\frac{8 k (k+3)}{u-(k+5)} + (k+4) \left((k+1) (k+5) u -k^3-3 k^2-29 k-3\right) \Bigg)
     }\,,
\end{align}
where $ d_k^{[1]} = 2 (k+3) \Gamma (k+4)$ and $u=2k+14-s-t$.

For Sequence 2, in the same way, we obtain the correlator $\langle \mathcal{O}_2 \bar{\mathcal{O}}_3 \bar{\mathcal{D}}_{k} \mathcal{D}_{k+1}  \rangle$ in Mellin space,
\begin{align}
\mathcal{M}_{2,3,k+4,k+5} \,=\; &{}
\scalemath{0.95}{
\frac{1}{d_k^{[2]}}\Biggl(\frac{(k+4)(k+5) u^2 - 2 (k+4)\left(
k^2+8k+27\right) u+k^4+15 k^3+106 k^2+384 k+592}{s-3}
}
 \nonumber\\ \label{eq:23kkp1-sugra}
    &{}\qquad\qquad 
    \scalemath{0.95}{
    +\frac{8 k (k+1)}{u-(k+4)}+ (k+4) \Big((k+5) u-k^2-3k-22\Big) \Biggr) 
    }
    \;,
\end{align}	
where $d_k^{[2]}=2\Gamma(k+5)$ and, as before, $u=2k+14-s-t$.

\subsection{Superconformal Ward identities}

Each of the two sequences of all-light correlators of two CPOs and two descendants given above is related by a superconformal Ward identity to a  sequence of correlators of four CPOs in the supergravity regime, in which the descendant $\cD_q$ is replaced by the CPO $\cO_{q+2}$. 
We next analyze these Ward identities. This will enable us to confirm the expression for the sequence $\langle \cO_3 \bar\cO_3 \bar\cO_{p} \cO_{p}\rangle$ derived in~\cite{Aprile:2017xsp}, using the Mellin space formula of~\cite{Rastelli:2016nze,Rastelli:2017udc}.
It will also enable us to derive an explicit compact expression for the dynamical part of the sequence $\langle \cO_2 \bar\cO_3 \bar\cO_{p} \cO_{p+1}\rangle$, and to verify that this expression agrees with the results of~\cite{Rastelli:2016nze,Rastelli:2017udc,Arutyunov:2018neq,Arutyunov:2018tvn,Caron-Huot:2018kta}.

For the correlators involving two CPOs and two descendants, for Sequence 1 and 2 respectively, from \eqref{eq:cK-defn} we have
\be
\mathcal{C}^{(\mathrm{\sst{desc}})}_{3,3,\:\! k+4,\:\!k+4}(U,V) \,=\, \frac{1}{U^{k+4}} \mathcal{G}^{(\mathrm{\sst{desc}})}_{3,3,\:\!k+4,\:\!k+4}(U,V)  \,, 
     \qquad
     \mathcal{C}^{(\mathrm{\sst{desc}})}_{2,3,\:\!k+4,\:\!k+5}(U,V) \,=\, \frac{1}{U^{k+5}} \mathcal{G}^{(\mathrm{\sst{desc}})}_{2,3,\:\!k+4,\:\!k+5}(U,V)  \,.
\ee
Similarly, for the corresponding correlators of four CPOs, we have 
\be
\mathcal{C}^{(\mathrm{\sst{CPO}})}_{3,3,\:\!k+2,\:\!k+2}(U,V) \,=\, \frac{1}{U^{k+2}} \mathcal{G}^{(\mathrm{\sst{CPO}})}_{3,3,\:\!k+2,\:\!k+2}(U,V)  \,, \qquad
     \mathcal{C}^{(\mathrm{\sst{CPO}})}_{2,3,\:\!k+2,\:\!k+3}(U,V) \,=\, \frac{1}{U^{k+3}} \mathcal{G}^{(\mathrm{\sst{CPO}})}_{2,3,\:\!k+2,\:\!k+3}(U,V)  \,.
\ee

In order to compare with results in the field theory literature, we introduce a complex null six-dimensional vector $\vec{y}$, such that the general single-trace CPO can be written in the form~\cite{Arutyunov:2002fh,Nirschl:2004pa,Dolan:2006ec}
\be
    \cO_p(\vec{y}) \,=\, y^{i_1} \ldots y^{i_p} \, \mathrm{Tr} \left( \phi_{i_1} \ldots \phi_{i_p} \right)\,, \qquad \vec{y}\cdot \vec{y} \,=\, 0\,.
\ee
We also introduce the cross-ratio-like variables
\be
    \sigma \,=\, \frac{y_{13}^2 y_{24}^2}{y_{12}^2 y_{34}^2} \:, \qquad \tau \,=\, \frac{y_{14}^2 y_{23}^2}{y_{12}^2 y_{34}^2} \:, \qquad
    y_{ij}^2 \,\equiv\, \vec{y}_i \cdot \vec{y}_j \:.
\ee
In terms of these variables, the stripped correlators $\cG$ take the form~\cite{Nirschl:2004pa} 
\be
   {\mathcal{G}}^{(\mathrm{\sst{CPO}})}_{p_1 p_2 p_3 p_4}(U,V,\sigma,\tau) \;=\; \mathcal{I}(U,V,\sigma,\tau) {\mathcal{H}}^{(\mathrm{\sst{CPO}})}_{p_1 p_2 p_3 p_4}(U,V,\sigma,\tau)\,,
\ee
where
\be
   \mathcal{I}(U,V,\sigma,\tau) \,=\, V + \sigma ^2 U V + \tau ^2 U + \sigma  V (V-U-1) +\tau  (1-U-V)  + \sigma  \tau  U (1-U-V)\,. 
\ee
In our conventions, the supergravity computation gives rise to the sequences of correlators $\langle \cO_3 \bar\cO_3 \bar\cO_{p} \cO_{p}\rangle$ and $\langle \cO_2 \bar\cO_3 \bar\cO_{p} \cO_{p+1}\rangle$ with $\mathcal{O}_p \sim \mathrm{Tr} \, Z^p$ and $\,Z = \phi_1+i\phi_2$. So the polarization vectors $\vec{y}$ are
\be
   \vec{y}_1 \,=\, \vec{y}_4 \,=\, \left(1, i,\vec{0}\right)\,, \qquad \vec{y}_2 \,=\, \vec{y}_3 \,=\, \left(1,-i,\vec{0}\right)\,,
   \quad~\Rightarrow\quad~ \sigma \,=\,1\,, \quad \tau \,=\, 0
    \,,
\ee
and therefore $\mathcal{I}=V^2$.

Let us now discuss the superconformal Ward identity that relates the correlators of two CPOs and two descendants to those of four CPOs. For the $22pp$ sequence of correlators, it was recently found~\cite{Turton:2024afd} that the Ward identity involves the differential operator (which appeared before in~\cite{Eden:2000bk,Drummond:2006by,Goncalves:2014ffa})
\be
\label{eq:WI-diff-op}
    \Delta^{(2)} \,=\, U\partial_U^2 + V\partial_V^2 + \left(U+V-1\right)\partial_U\partial_V + 2\left(\partial_U+\partial_V\right)\,,
\ee
and takes the form
\be
\label{eq:WI}
    \mathcal{C}^{(\mathrm{\sst{desc}})}_{\{\Delta_i\}}(U,V) \,=\, \left(\Delta^{(2)}\right)^2 {\mathcal{C}}^{(\mathrm{\sst{CPO}})}_{\{\Delta_i\}}(U,V) \:.
\ee
For $p=2$, this was based on previous analyses of the relation of correlators of four descendants $\mathcal{D}_0$ (Lagrangian-density operators) to correlators of four CPOs of dimension two~\cite{Drummond:2006by,Goncalves:2014ffa}. {\it A priori}, for the remainder of the $22pp$ sequence, the Ward identity could have depended on $p$, however it was found by explicit computation and matching to known results that \eqref{eq:WI-diff-op}, \eqref{eq:WI} were sufficient.

For the two sequences of correlators computed in the present work, we will similarly make the working assumption that \eqref{eq:WI-diff-op}, \eqref{eq:WI} will remain sufficient to relate the correlators of two CPOs and two descendants to the correlators of four CPOs. We shall not attempt a first-principles derivation of this assumption here; this is an interesting question that could be investigated using the results of~\cite{Caron-Huot:2018kta}, and that we leave for future work. Rather, this will be a prescription that works in practice, that will be justified by comparison to the results of~\cite{Rastelli:2016nze,Rastelli:2017udc,Arutyunov:2018neq,Arutyunov:2018tvn,Aprile:2017xsp}.

Let us first discuss Sequence 1.
Using the Mellin space conjecture of~\cite{Rastelli:2016nze,Rastelli:2017udc}, an expression for the dynamical part of the 33$pp$ CPO correlator in the supergravity regime was derived in~\cite{Aprile:2017xsp}, for $p\ge 3$. Upon specializing to  $\sigma=1$, $\tau =0$, this becomes (for the definition of the $\bar{D}$-functions, see Eq.~\eqref{eq:Dbar-def})
\be
\label{eq:33pp-CPO}
    {\mathcal{H}}^{(\mathrm{\sst{CPO}})}_{3 3 p p} \:=\: U^p \, \left( \bar{D}_{p-1,\, p+2,\, 2,\, 3}  +\frac{1}{p-2}\bar{D}_{p,\, p+2,\, 2,\, 2} +\frac{p-1}{p-2}\bar{D}_{p,\, p+2,\, 3,\, 3}  \right).  
\ee
First, we act on~\eqref{eq:33pp-CPO} with the above differential operator to evaluate the right-hand side of the Ward identity~\eqref{eq:WI}. We then simplify the answer using the following differential properties of the $\bar{D}$-functions~\cite{Gary:2009ae},
\begin{align}
\begin{aligned}
\partial_U \bar{D}_{\Delta_1,\Delta_2,\Delta_3,\Delta_4}(U,V) \;&=\;    
-\bar{D}_{\Delta_1+1,\Delta_2+1,\Delta_3,\Delta_4}(U,V) \,, \\
\partial_V \bar{D}_{\Delta_1,\Delta_2,\Delta_3,\Delta_4}(U,V) \;&=\;    
-\bar{D}_{\Delta_1,\Delta_2+1,\Delta_3+1,\Delta_4}(U,V) \,.
\end{aligned}
\end{align}
Next, we transform to Mellin space. We compare the resulting expression to our supergravity result~\eqref{eq:33kk-sugra} and use the residue of the pole at $s=2$ to fix the overall normalization. We then obtain precise agreement between the CFT conjecture~\eqref{eq:33pp-CPO} and our supergravity result for $\mathcal{M}_{33,k+4,k+4}$ in Eq.~\eqref{eq:33kk-sugra}, with $p=k+2$.

For Sequence 2, we start with one example of the sequence of correlators of CPOs, 
$\cC^{(\mathrm{\sst{CPO}})}_{2,3,p,p+1}$,
namely $p=4$, which was computed in~\cite{Arutyunov:2018neq} via Witten diagrams. For this case, translating the expression of~\cite{Arutyunov:2018neq} into our notation, the $\mathcal{H}$ function takes the form
\be       
{\mathcal{H}}^{(\mathrm{\sst{CPO}})}_{2, 3, 4,5} 
\:=\:
\frac{U^2}{V} \, \bar{D}_{2,3,4,7}  \;.  
\ee
We follow the same steps as done above for Sequence 1, i.e.~we evaluate the right-hand side of the Ward identity and then convert to Mellin space. This time, we use the residue of the pole at $s=3$ to fix the overall normalization. After doing so, we obtain precise agreement with~\eqref{eq:23kkp1-sugra} for $k=2$.

Based on this agreement, we propose the following expression for the 
$\mathcal{H}$ function of the full sequence  $\cC^{(\mathrm{\sst{CPO}})}_{2,3,p,p+1}$ of correlators of CPOs,
\be      
\label{eq:H23ppp1-new}
{\mathcal{H}}^{(\mathrm{\sst{CPO}})}_{2, 3,\:\! p,\:\! p+1} \:=\: \frac{U^2}{V} \, \bar{D}_{2,3,\:\! p,\:\! p+3}  \;.
\ee
Carrying out the same steps as for $p=4$, we obtain precise agreement with our supergravity result for  $\mathcal{M}_{2,3,k+4,k+5}$,~\eqref{eq:23kkp1-sugra}, for all $k$, again with $p=k+2$.

We have checked explicitly that the expression~\eqref{eq:H23ppp1-new} is consistent with the position-space results of~\cite{Arutyunov:2018tvn} over the finite range of overlap, $2 \le p \le 7$. 
Moreover, our expression for $\cH_{2,3,p,p+1}$ is also in agreement with the Mellin space formula of~\cite{Rastelli:2016nze,Rastelli:2017udc}. To show this, we reorder the operators such that $p_1\geq p_2 \geq p_3 \geq p_4$, obtaining
\be
    \cH_{p+1, p, 3, 2}(U,V) \,=\, \frac{1}{U} \cH_{2, 3, p, p+1}(U,V) \,=\, \frac{U}{V} \bar{D}_{2, 3, p, p+3}(U,V)\,. 
\ee
Next, we take the appropriate Mellin transform for $\cH_{p+1, p, 3, 2}$ in $s$, $t$ and $\tilde{u}=u-4$ variables as defined in~\cite{Rastelli:2016nze,Rastelli:2017udc}, such that $s+t+\tilde{u}=2p+2$. Upon doing so, we obtain
\be
\label{eq:Mtpp1p32}
    \widetilde{M}_{p+1, p, 3, 2}(s,t,\tilde{u},\sigma,\tau) \;\propto\;  \frac{1}{(s-3)(t-p-1)(\tilde{u}-p)}\,. 
\ee
This can now be compared to the Mellin space formula in~\cite[Eq.\;(4.47)]{Rastelli:2017udc}. In that formula, for these correlators there is only one term in the sum\footnote{In~\cite[Eq.\;(4.47)]{Rastelli:2017udc}, for these correlators we have $L=2$, so $i=j=k=0$.}, and thus both $\widetilde{M}$ and $\cH$ are independent of $\sigma,\tau$. 
Evaluating this term, we find precise agreement with \eqref{eq:Mtpp1p32} for general $p$. 

Finally, we have also checked that our expression for $\cH_{2,3,p,p+1}$ is consistent with the generating function of~\cite{Caron-Huot:2018kta}, being, in effect, a particular Taylor expansion starting from the $\langle 2222 \rangle$ correlator of CPOs.
These precise agreements between the explicit compact expression~\eqref{eq:H23ppp1-new} and previous works justify {\it a posteriori} our prescription for the Ward identity, and give additional evidence in support of the expressions of~\cite{Rastelli:2016nze,Rastelli:2017udc,Caron-Huot:2018kta} for tree-level four-point correlators in the supergravity regime.



\section{Discussion}
\label{sec:discussion}

In this paper we have described the derivation of a new closed-form perturbative LLM solution, composed of a pair of linearized supergravitons of different mode numbers and their quadratic backreaction. We constructed the diffeomorphism that converts this solution into de Donder-Lorentz gauge. The resulting solution is the first such closed-form multi-mode solution, and is analogous to the large families of multi-mode solutions with AdS$_3 \times$S$^3$ asymptotics that have been constructed in recent years.

We used this solution to compute two sequences of tree-level four-point correlators of two single-particle CPOs and two descendants in the supergravity regime. These correlators are related via a superconformal Ward identity to tree-level correlators of four single-particle CPOs in the supergravity regime. For the first sequence, there is a proposed explicit formula for $\langle \cO_3 \bar{\cO}_3 \bar{\cO}_p \cO_p \rangle$ obtained from Mellin space methods~\cite{Aprile:2017xsp,Rastelli:2016nze,Rastelli:2017udc}, given in Eq.~\eqref{eq:33pp-CPO}, and our results are the first confirmation of this formula for general $p$ from a supergravity calculation.

From the second sequence of correlators of two CPOs and two descendants, we obtained a new compact explicit expression for $\langle \cO_2 \bar{\cO}_3 \bar{\cO}_p \cO_{p+1} \rangle$ for general $p$. We verified that it agrees with the results obtained from tree-level Witten diagrams in~\cite{Arutyunov:2018neq,Arutyunov:2018tvn} in the finite range of overlap, $2 \le p \le 7$. 
We also verified that it agrees with the Mellin space bootstrap conjecture of~\cite{Rastelli:2016nze,Rastelli:2017udc} for general $p$, as well as the generating function proposed in~\cite{Caron-Huot:2018kta}.

Our results can be regarded as further confirmation of the validity of the method of deriving all-light correlators from probing smooth supergravity solutions that are composed of backreacted supergravitons on the vacuum.
They can also be regarded as highly non-trivial checks of the Mellin space bootstrap conjecture of~\cite{Rastelli:2016nze,Rastelli:2017udc}, the Witten diagram computations of~\cite{Arutyunov:2018neq,Arutyunov:2018tvn},
the generating function of~\cite{Caron-Huot:2018kta}, 
and indeed of this instance of  AdS$_5$/CFT$_4$ holography.

Our results open up several possibilities for future work. Our method can be straightforwardly generalised to construct two-mode LLM solutions for any pair of modes with specific choices of mode numbers $m$ and $n$, and to compute other infinite sequences of four-point correlators of single-trace operators from such backgrounds. As remarked in the Introduction, this would allow access to correlators with an arbitrary degree of extremality. 
We took a first step in this direction by solving the linear-order equation for the probe for general $m$ and $n$.

Second, by extending our approach to higher orders in the perturbative expansion, it should be possible to compute four-point functions with more general multi-trace operators than powers of $\cO_2$, for instance the double-trace operators $\cO_3^2$ and $(\cO_2 \cO_3)$. This possibility to access more general multi-traces is an advantage that our ten-dimensional approach has over backgrounds that lie in a consistent truncation, that can access a relatively limited number of multi-trace operators. The disadvantage is that a ten-dimensional approach is more difficult to carry out in practice. Nevertheless, our results can also be regarded as laying the foundations for such investigations, and work in this direction is in progress.

A third potential direction for the future is to analyse the two-mode solution we have constructed in the present work using precision holographic heavy-light three-point correlators, generalizing the works of~\cite{Skenderis:2007yb,Turton:2025svk}.
This would require expanding the holographic  dictionary to operators of dimension higher than four.
It would also be interesting to compute holographic four-point correlators on asymptotically AdS$_5 \times $S$^5$ backgrounds with less supersymmetry, building on the works of~\cite{Chen:2007du,Lunin:2008tf,Jia:2023iaa,Ganchev:2025dzn}.

There remain many exciting questions to be explored about the dynamics of strongly coupled $\cN=4$ SYM, and of other holographic field theories. 
Smooth supergravity solutions that are holographically dual to 
heavy pure states in the field theory are important entries in the holographic dictionary that promise to yield many more valuable results in the future.


\section*{Acknowledgements}

For fruitful discussions, we thank Stefano Giusto, Rodolfo Russo, and Kostas Skenderis. 
The work of DT was supported by a Royal Society Tata University Research Fellowship and by the STFC Grant `New Frontiers in Particle Physics, Cosmology and Gravity', ST/X000583/1. The work of A.T.~was supported by a Royal Society Research Fellows Enhanced Research Expenses grant.


\begin{appendix}

\section{Diffeomorphism for the quadratic-order metric}
\label{app:metric-quad-diffeo}

In this appendix we give the explicit generator $\xi^{(2)}$ of the gauge transformation that removes singular terms involving negative powers of $\Sigma$ and converts the second-order metric for $n=2$, $m=3$ into De Donder-Lorentz gauge.
We separate and factor out the $\alpha_1$, $\alpha_2$ dependence as
\be
    \xi^{(2)} \,=\, \alpha_1^2  \;\! \xi_{(2,0)} + \alpha_1 \alpha_2  \;\! \xi_{(1,1)} +
    \alpha_2^2  \;\! \xi_{(0,2)}    \;.
\ee
$\xi_{(2,0)}$ was given in~\cite{Turton:2024afd}, so we focus on $\xi_{(0,2)}$ and $\xi_{(1,1)}$.
The components of $\xi_{(0,2)}$ are
\be
    \xi_{(0,2)}^{t} \,=\, -\frac{\left(1-6 R^2+\cos2 \theta\right) \sin^6\theta  \sin 6(t-\phi)}{16 \left(R^2+1\right)^4 \Sigma }\;,
\ee

\begin{align}
    \xi_{(0,2)}^{R} &\,=\, -\frac{1}{128 \left(R^2+1\right)^3 \Sigma ^2} R\Bigg( 16\left(R^2+1\right) \left(5 R^2-3\right) \cr
    &{}\qquad\qquad\qquad\qquad\qquad\qquad +\left(11+24 R^4+12 \left(4 R^2+1\right) \cos2 \theta+\cos4 \theta\right) \sin^6\theta \cos 6 (t-\phi )\Bigg) \cr 
     &{}\qquad
    +\frac{1}{64 \left(R^2+1\right) \Sigma ^3} R \Bigg(-4 \left(3 R^6+8 R^4+6 R^2+1\right) \cos2 \theta  \cr
    &{}\qquad\qquad\qquad\qquad\qquad\qquad +\left(1+R^2\right) \left(-6+R^2+4 R^4+\left(3 R^2+2\right) \cos 4 \theta\right)-48 R^2 \sin^6\theta\Bigg) \cr
    &{}\qquad + \frac{R \left(3-8 R^2\right) \sin^6\theta}{8 \left(R^2+1\right)^3 \Sigma }+\frac{R \left(7 R^2-2\right) \sin^6\theta}{8 \left(R^2+1\right)^4} 
     +\frac{3 R  \sin^6\theta \cos 6( t-\phi )}{16 \left(R^2+1\right)^3} \\
    &{}\qquad + \frac{1}{8960 \left(R^2+1\right)^4} R \Bigg(\left(3672+35R^2 \left(48 R^4+320 R^2+491\right)\right) \cos 2 \theta \cr
    &{}\qquad\qquad\qquad\qquad\qquad\qquad
    -42 \left(20 R^4+55 R^2+14\right) \cos 4 \theta + 7 \left(25 R^2+8\right)  \cos 6 \theta\Bigg)\,, \nonumber
\end{align}

\begin{align}
     \xi_{(0,2)}^{\theta} &\,=\,-\frac{\sin \theta  \cos^3\theta}{2\Sigma ^2} -\frac{\left(1- 6 R^2+\cos 2 \theta\right) \cos\theta\sin^7\theta  \cos 6 (t-\phi)}{16 \left(R^2+1\right)^3\Sigma ^2} \cr
   &{}\qquad + \frac{1}{6720 \left(R^2+1\right)^4}\Bigg(5 \left(84 R^4+161 R^2+32\right)\\ 
   &{}\qquad\qquad\qquad\qquad +14 \left(30 R^4+50 R^2+11\right) \cos2\theta-21 \left(5 R^2+2\right) \cos4\theta\Bigg) \sin2\theta\,.\nonumber
\end{align}

The components of $\xi_{(1,1)}$ that involve the mode $m-n=1$ are as follows. (As described in the main text, we disregard the terms involving the mode $m+n=5$.)
\be
   \xi_{(1,1)}^{\phi} \,=\, \frac{ \left(25 \sin \theta \left(-7 \cos 2\theta+28 R^2+73\right)-2 \left(35 R^2+109\right) \csc\theta\right)\sin (t-\phi )}{2800 \left(R^2+1\right)^{7/2}}\,,
\ee

\begin{align}
   \xi_{(1,1)}^{t} &\,=\,\frac{1}{512 \left(R^2+1\right)^{9/2} \Sigma} \Bigg(\left(64 R^8+816 R^6+2768 R^4+829 R^2+38\right) \sin\theta \cr 
   &{}\qquad\qquad\qquad\qquad\qquad  -\left(272 R^6+1264 R^4+283 R^2+26\right) \sin3\theta \cr
   &{}\qquad\qquad\qquad\qquad\qquad -\left(-128 R^4+167 R^2+50\right) \sin5\theta + 7 \left(7 R^2+2\right) \sin7 \theta \Bigg) \sin (t-\phi ) 
   \cr
   &{}
   + \frac{1}{11200 \left(R^2+1\right)^{9/2}} \Bigg(-7 \left(960 R^4+3760 R^2+943\right) \\
   &{}
   \qquad\qquad\qquad\qquad\qquad 
   + 200 \left(56 R^4+223 R^2+52\right) \cos2 \theta \cr 
   &{}
   \qquad\qquad\qquad\qquad\qquad -105 \left(80 R^2+23\right) \cos4 \theta  
   \Bigg) \sin\theta \sin (t-\phi )\,, \nonumber
\end{align}

\begin{align}
     \xi_{(1,1)}^{R} &\,=\,\frac{1}{1024 \left(R^2+1\right)^{7/2} \Sigma ^2} R \Bigg(-396 -1477 R^2+32 \left(4 R^6+18 R^4-R^2-63\right) R^4 \cr 
      &{}\qquad\qquad\qquad\qquad\qquad\qquad
      -8 \left(88 R^8+208 R^6+132 R^4+145 R^2+42\right) \cos2\theta  \cr
      &{}\qquad\qquad\qquad\qquad\qquad\qquad  +4 \left(63 R^2-24 \left(R^2+3\right) R^4+20\right) \cos4 \theta \cr
      &{}\qquad\qquad\qquad\qquad
      +8 \left(4 R^4-7 R^2+2\right) \cos6 \theta  + \left(9 R^2-4\right) \cos 8 \theta \Bigg) \sin\theta \cos (t-\phi ) \cr 
       &{}
       +\frac{1}{11200 \left(R^2+1\right)^{9/2}} R \Bigg( 7\left(-1763-340 R^2+200 \left(R^2+3\right) R^4\right) \\
   &{}~~
   \qquad\qquad\qquad\qquad\qquad
   -200 \left(42 R^4+77 R^2+9\right) \cos2 \theta \cr
   &{}~~
   \qquad\qquad\qquad\qquad\qquad + 525 \left(4 R^2+1\right) \cos4\theta
   \Bigg) \sin\theta \cos (t-\phi ) 
   \,, \nonumber
\end{align}

\begin{align}
     \xi_{(1,1)}^{\theta} \,=&\; \frac{1}{256 \left(R^2+1\right)^{7/2} \Sigma ^2} \Bigg(-60-126 R^2+16 \left(2 R^4+5 R^2+2\right) R^4 \cr 
      &{}\qquad\qquad\qquad\qquad 
      -\left(46+131 R^2+80 \left(R^2+3\right) R^4\right) \cos 2\theta \\
       &{}\qquad\qquad\qquad\qquad 
       -2 \left(-8 R^4+R^2-6\right) \cos4\theta -\left(2-3 R^2\right) \cos 6\theta \Bigg) \sin^2\theta\cos\theta  \cos (t-\phi) \cr
       &{}
       -\frac{ \left(4547+1680 R^2-1400 \left(R^2+3\right) \cos 2 \theta+525 \cos 4 \theta  \right) \cos\theta \cos (t-\phi )}{11200 \left(R^2+1\right)^{7/2}}\;. \nonumber
\end{align}


\section{Solution to the linearised equation of motion}

\label{app:recur-sol}

In this appendix, we derive the solution \eqref{eq:phi-i-sol} to the linear order equation of motion \eqref{eq:o-a-eom-n-g} for $k \ge 2$, which we reproduce here for convenience:
\be
\label{eq:o-a-eom-n-g-app}
    \square^{(0)} \Phi^{(1)'} \,=\, \frac{16 n (n-1) k (k-1)}{n+1} s_n \hat{B}_{k+4} Y^{(k,2-k)} Y_{n-2}\,. 
\ee
We recall that the zeroth order background metric on $S^5$ is given by
\be
    ds^2 \,=\, d\theta^2+\sin^2\theta d\phi^2 +\cos^2\theta d\Omega_3^2\,.
\ee
We use the following definition of the $SO(4)$-invariant spherical harmonics,  
\be
\label{eq:sph-harm-def-1}
    Y^{(k,m)} = y^k_m(\theta) e^{i m \phi}\,, 
\ee
where 
\be
\label{eq:sph-harm-def-2}
    y^k_m(\theta) \,=\, x^{|m|} {}_2 F_1\left(-\frac{k}{2}+\frac{|m|}{2},2+\frac{k}{2}+\frac{|m|}{2},1+|m|;x^2\right)\,, \qquad x \,=\, \sin\theta\,.
\ee

We first consider the generic case, $k \ge n \ge 2$. In this case, we observe the following identity,
\be
    \square_{\mathrm{\sst S}} \left( Y^{(k,2-k)} Y_{n-2} \right) \,=\, \left(4-(k+n)^2\right) Y^{(k,2-k)} Y_{n-2} + 4(n-2)(k-2) Y^{(k-1,3-k)} Y_{n-3}\,,   
\ee
which can be generalised by relabelling $n\to n-i$ and  $k\to k-i$ to obtain, for $0 \le i \le n-2$,
\begin{align}
    \square_{\mathrm{\sst S}} \left( Y^{(k-i,2+i-k)} Y_{n-i-2} \right) \,&=\, \left(4-(k+n-2i)^2\right) Y^{(k-i,2+i-k)} Y_{n-i-2} \cr
    &\qquad {}+ 4(n-i-2)(k-i-2) Y^{(k-i-1,3+i-k)} Y_{n-i-3}\,.
\end{align}
Based on this set of identities and the source term of Eq.~\eqref{eq:o-a-eom-n-g-app}, we make the following Ansatz for $\Phi^{(1)'}$, 
\be
    \Phi^{(1)'} \,=\, \sum_{i=0}^{n-2} \Phi_{(i)} Y^{(k-i,2+i-k)} Y_{n-i-2}\;.  
\ee
This gives rise to the following recursive system of equations, 
\begin{align}
    \square_{\mathrm{\sst A}} \Phi_{(0)} + \left( 4-(n+k)^2\right) \Phi_{(0)} &\,=\, \frac{16 n (n-1) k (k-1)}{n+1} s_n \hat{B}_{k+4}\,, \\
    \square_{\mathrm{\sst A}} \Phi_{(i)} + \left(4-(k+n-2i)^2\right)\Phi_{(i)} &\,=\, - 4(n-i-1)(k-i-1) \Phi_{(i-1)}\,, \quad i=1,\ldots,n-2\,.
\end{align}
In principle, the solution to these equations should be added to the solution of the corresponding homogeneous equations, which are
\be
  \Phi_{(i)} \,=\, \hat{B}_{k+n+2-2i}\;.   
\ee
However, we shall drop this part of the solution in what follows, since it does not contribute to the dynamical part of the four-point correlators we study.  

We also notice the following identity,
\be
    \square_{\mathrm{\sst A}}\left(s_{n-1}\hat{B}_{k+3}\right) + \left(4-(k+n)^2\right) s_{n-1}\hat{B}_{k+3} \;=\;  -\frac{4 n^2 (k+3)}{n+1} s_n \hat{B}_{k+4}\;. 
\ee
This yields the solution of the first equation of the recursive system,
\be
    \Phi_{(0)} \,=\, - \frac{4 (n-1) k(k-1)}{n (k+3)} s_{n-1}\hat{B}_{k+3}\,. 
\ee
For $n=2$, this gives
\be
    \Phi_{(0)} \,=\, -\frac{2 k(k-1)}{k+3} s_{1}\hat{B}_{k+3}\,,
\ee
and therefore~\cite{Turton:2024afd}
\be
\label{eq:phi1p-n2-app}
    \Phi^{(1)'} \,=\, -\frac{2 k(k-1)}{k+3} s_{1}\hat{B}_{k+3} Y^{(k,2-k)}\,, 
\ee
as given in \eqref{eq:phi1pkge2-tt1} of the main text. 
Similarly, for $k\ge n \ge 3$, one can find a solution for $\Phi_{(1)}$,
\be
    \Phi_{(1)} \,=\, - \frac{4(n-2) k (k-1)(k-2)}{(n-1)(k+3)(k+2)} s_{n-2} \hat{B}_{k+2}\,. 
\ee

To solve for general $i$, we can use the identity 
\be
    \square_{\mathrm{\sst A}}\left(s_{n-i-1}\hat{B}_{k-i+3}\right) + \left(4-(k+n-2i)^2\right) s_{n-i-1}\hat{B}_{k-i+3} \,=\,  -\frac{4 (n-i)^2 (k-i+3)}{n-i+1} s_{n-i} \hat{B}_{k-i+4}\,,
\ee
whereupon we obtain the solution $\Phi^{(1)'}$ for general $k\ge n \ge 2$,
\be
\label{eq:phi-i-sol-app}
   \Phi_{(i)} \,=\, -\frac{4(n-i-1)}{n-i}\frac{(k-i+2)(k-i+1)(k-i)(k-i-1)}{(k+3)(k+2)(k+1)} s_{n-i-1} \hat{B}_{k-i+3} \,, \quad i=0,\ldots,n-2\,,
\ee
as given in \eqref{eq:phi-i-sol} of the main text. (Note that we have included $i=0,1$ in \eqref{eq:phi-i-sol-app}).

Next, the solutions for $k<n$ can be derived from careful limits of the solution for $k\ge n$, as follows.  Evaluating the spherical harmonics as defined in Eqs.\;\eqref{eq:sph-harm-def-1}, \eqref{eq:sph-harm-def-2}, for general $k\ge 2$, we observe the identity 
\be
   (k-1) Y^{(k,2-k)} \,=\, \left(k-1 -( k+1)\sin^2\theta\right) e^{i \phi  (2-k)} (\sin \theta)^{k-2}\,,  \qquad k \ge 2 \,,\label{eq:id1}
\ee
and more generally, shifting $k \to k-i$, for $k\ge i+2$ we have
\be
   (k-i-1) Y^{(k-i,2+i-k)} \,=\, \left(k-i-1 -( k-i+1)\sin^2\theta\right) e^{i \phi  (2+i-k)} (\sin \theta)^{k-i-2}\,,  \qquad k \ge i+2 \,.\label{eq:id-gen}
\ee
We start by evaluating the above solution for $k\ge n$ to obtain expressions that involve the right-hand sides of \eqref{eq:id1} and \eqref{eq:id-gen}. Then the desired limits for any particular $k<n$ can be taken. 
For instance, for $n=2$, from the $k\ge 2$ solution \eqref{eq:phi1p-n2-app}, to obtain the solution for $k=1$, we use the identity \eqref{eq:id1} and then set $k=1$, obtaining~\cite{Turton:2024afd}
\be
\Phi^{(1)'}_{k=1,n=2} \,=\, \hat{B}_{4}(x) s_{1}(x) Y_{1}(y)\,.
\ee
For $n=3$, evaluating \eqref{eq:phi-i-sol-app}, the solution for $k\ge 3$ is
\be
\Phi^{(1)'}_{k\ge 3} \,=\, -\frac{8}{3}\frac{k(k-1)}{k+3} s_{2}\hat{B}_{k+3} Y^{(k,2-k)} Y_1-\frac{2k(k-1)(k-2)}{(k+3)(k+2)} s_1 \hat{B}_{k+2} Y^{(k-1,3-k)} \,.
\ee
Using the identities \eqref{eq:id1} and \eqref{eq:id-gen} for $i=1$, we similarly find that when $k=1$ only the first term survives, and when $k=2$, both terms surivive:
\begin{align}
\begin{aligned}
\Phi^{(1)'}_{k=1,n=3} \,&=\, \frac{4}{3} s_2(x) \hat{B}_4(x) Y_2(y) \,, \label{eq:phi1p-tt3-app} \\[3mm]
\Phi^{(1)'}_{k=2,n=3} \,&=\, -\frac{16}{15} s_2(x) \hat{B}_5(x) Y^{(2,0)}(y) Y_1(y) + \frac{2}{5} s_1(x) \hat{B}_4(x) Y_1(y) \,,  \\[2mm]
\end{aligned}
\end{align}
where the solution for $k=2$, $n=3$ is given in the main text in Eq.\;\eqref{eq:phi1p-tt3-new}.

For general $n$, the solution \eqref{eq:phi-i-sol-app} for $\Phi^{(1)'}$ for general $k\ge n$ consists of a sum of $n-1$ terms. 
Again, for $k=1$, only one term survives, which for general $n$ is given by
\be
    \Phi^{(1)'}_{k=1} \,=\, \frac{2(n-1)}{n} s_{n-1}(x) \hat{B}_4(x) Y_{n-1}(y)\,,
\ee
as given in the main text in Eq.\;\eqref{eq:k1sol}. 
In general, for $1\le k < n$, there are $k$ terms that survive.


\section{Sources of the quadratic equation of motion}

\label{app:sources-both}

In this appendix we record the part of the sources of the projected quadratic order equations of motion that contribute to the dynamical part of the two sequences of four-point correlators. We write them in the form that arises after rewriting as sums of products of one of each of the functions $B^{\pm}_\Delta$, $\hat{B}_\Delta$ defined in Eq.~\eqref{eq:b-to-bdy-hy}. For ease of presentation, we suppress the spacetime dependence. 

For Sequence 2, the source is given by
\begin{align}
   \!\!\!
   \mathcal{J}^{[2]}
   \;=&\;\frac{2}{5} (k-1) k B_2^{-} B_2^{+}\hat{B}_{k+5} - \frac{2}{5} (k-6) (k+5) B_3^{-} B_2^{+} \hat{B}_{k+6} + \frac{4}{5} k (7 k+23) B_1^{-} B_3^{+} \hat{B}_{k+5} \nonumber\\ 
   & - \frac{8}{5} (k+5) (2 k+3) B_2^{-} B_3^{+}\hat{B}_{k+6} - \frac{3}{7} \left(7 k^3+54 k^2+109 k+28\right) B_4^{-} B_3^{+}\hat{B}_{k+6} \nonumber\\
   & + \frac{\left(105 k^4+1475 k^3+6768 k^2+11252 k+3920\right) }{35 (k+5)} B_3^{-} B_3^{+}\hat{B}_{k+5}  + \frac{24 k (k+3)}{k+4}  B_1^{-} B_4^{+}\hat{B}_{k+4}\nonumber\\
   & - \frac{12}{5} (k-1) (k+5) B_1^{-} B_4^{+}\hat{B}_{k+6}  + \frac{6 \left(47 k^3+510 k^2+1858 k+2100\right) }{35 (k+5)} B_2^{-} B_4^{+}\hat{B}_{k+5}\nonumber\\
   &  + \frac{24 k}{k+4}  B_2^{-} B_4^{+}\hat{B}_{k+3} + \frac{3 \left(317 k^4+4447 k^3+22268 k^2+52068 k+45640\right) }{175 (k+4) (k+5)} B_3^{-} B_4^{+} \hat{B}_{k+4}\nonumber\\
   & +\left(-7 k^3-\frac{2643 k^2}{35}-\frac{8392 k}{35}-216\right)  B_3^{-} B_4^{+}\hat{B}_{k+6} + 3 (k+3) \left(k^2+8 k+12\right) B_4^{-} B_4^{+}\hat{B}_{k+7} \nonumber\\
   & - \frac{36}{35} \left(2 k^2+17 k+28\right) B_5^{-} B_4^{+}\hat{B}_{k+6}  -\frac{18}{35} \left(17 k^2+113 k+140\right) B_2^{-} B_5^{+} \hat{B}_{k+6} \nonumber\\
   & + 4 (k+3) \left(k^2+8 k+12\right) B_3^{-} B_5^{+}\hat{B}_{k+7} + \frac{24}{175} \left(127 k^2+698 k+1050\right) B_4^{-} B_5^{+}\hat{B}_{k+6} \nonumber\\
   & + \frac{4}{35} \left(109 k^2+776 k+1470\right) B_3^{-} B_6^{+}\hat{B}_{k+6} + \frac{12 \left(109 k^2+776 k+1470\right)}{35 (k+5)}  B_4^{-} B_6^{+}\hat{B}_{k+5} \nonumber\\
   & -\frac{48 k}{k+4} B_2^{-} B_5^{+}\hat{B}_{k+4} - 24 k B_1^{-} B_5^{+}\hat{B}_{k+5} -\frac{\left(5 k^3+43 k^2+262 k+200\right)}{5 (k+4)} B_2^{-} B_3^{+}\hat{B}_{k+4}  \nonumber\\
   & - \frac{144 \left(2 k^2+17 k+28\right) }{35 (k+5)}B_5^{-} B_5^{+}\hat{B}_{k+5} -\frac{6 \left(361 k^3+2280 k^2+3299 k-1260\right)}{175 (k+5)}   B_4^{-} B_4^{+}\hat{B}_{k+5}\nonumber\\
   & -\frac{12 \left(251 k^3+3097 k^2+13312 k+18200\right)}{175 (k+5)}  B_3^{-} B_5^{+}\hat{B}_{k+5} - \frac{48 k}{k^2+7 k+12}  B_3^{-} B_5^{+}\hat{B}_{k+3} \nonumber\\
   & -\frac{12 \left(317 k^3+2370 k^2+7243 k+7140\right)}{175 (k+4) (k+5)}  B_4^{-} B_5^{+}\hat{B}_{k+4} \,.
\end{align}

\newpage

For Sequence 1, the source is given by
\begin{adjustwidth}{-4mm}{-4mm} 
\scalebox{0.95}{%
\parbox{\linewidth}{%
\begin{align}
     \!\!\!
   \mathcal{J}^{[1]}
   \;=&\; \frac{1}{4} (k+3) (k+4)  B_1^{-} B_1^{+}\hat{B}_{k+4} + \frac{1}{4} (k-12) (k+4)  B_2^{-} B_1^{+}\hat{B}_{k+5} \nonumber \\
   &- \frac{(k-5) (k+4) (k+5)  }{2 (k+3)}B_3^{-} B_1^{+}\hat{B}_{k+6}+ \frac{3}{4} (k+4) (3 k-4)  B_1^{-} B_2^{+}\hat{B}_{k+5} \nonumber\\ 
   &  + \frac{\left(57 k^3+406 k^2+413 k-1964\right) }{8 (k+3)} B_3^{-} B_2^{+}\hat{B}_{k+5} + \frac{\left(105 k^3+614 k^2+477 k-1964\right)  }{8 (k+3)} B_2^{-} B_3^{+}\hat{B}_{k+5}\nonumber\\
   & + \frac{\left(128 k^5+2591 k^4+19072 k^3+63553 k^2+94968 k+66000\right)  }{32 (k+3) (k+5)} B_4^{-} B_3^{+}\hat{B}_{k+5} \nonumber\\
   & +\frac{24 k \left(k^2+k-2\right)  }{(k+3)^2}B_1^{-}B_4^{+}\hat{B}_{k+3}  + \frac{24 (k-1) k  }{(k+3)^2}B_2^{-} B_4^{+}\hat{B}_{k+2} + \frac{24 k (k+1)  }{k+3}B_2^{-} B_4^{+}\hat{B}_{k+4}
   \nonumber\\
   & + \frac{48 k (k+1)}{(k+3)^2} B_3^{-} B_4^{+} \hat{B}_{k+3}+\frac{\left(799 k^4+10368 k^3+46913 k^2+85368 k+66000\right) }{32 (k+3) (k+5)} B_3^{-} B_4^{+}\hat{B}_{k+5} \nonumber\\
   & + \frac{\left(-8473 k^4-71251 k^3-112811 k^2+1065979 k+2690076\right)  }{1120 (k+3) (k+5)}B_5^{-} B_4^{+}\hat{B}_{k+5} \nonumber\\
   & +4 \left(k^3+9 k^2+20 k+12\right) \left( B_5^{-} B_4^{+} + B_4^{-} B_5^{+} \right) \hat{B}_{k+7} \nonumber\\
   & +\frac{\left(4967 k^4+36269 k^3+115669 k^2+1200379 k+2690076\right)}{1120 (k+3) (k+5)}   B_4^{-} B_5^{+}\hat{B}_{k+5}\nonumber\\
   & +\frac{\left(863 k^4+14096 k^3+68641 k^2+202696 k+272784\right) }{70 (k+3) (k+4) (k+5)}  B_5^{-} B_5^{+}\hat{B}_{k+4}\nonumber\\
   & + \frac{\left(1937 k^3+9420 k^2-5153 k-51396\right)  }{140 (k+3)}B_5^{-} B_5^{+}\hat{B}_{k+6}-\frac{24 (k-1) k  }{k+3} B_1^{-} B_5^{+}\hat{B}_{k+4}
   \\
   & -\frac{24 k (k+1)}{k+3} B_2^{-} B_5^{+}\hat{B}_{k+5}  -\frac{(k+4) (k+5) (3 k+1)}{2 (k+3)}  B_2^{-} B_2^{+}\hat{B}_{k+6} \nonumber\\ 
   & -\frac{(k-5) (k+4) (k+5)}{2 (k+3)}  B_1^{-} B_3^{+}\hat{B}_{k+6}  - \frac{(k+4) (k+5) (21 k-17) }{2 (k+3)} B_3^{-} B_3^{+}\hat{B}_{k+6} \nonumber\\
   & -\frac{(k+4) \left(9 k^2+32 k-81\right) }{4 (k+3)}  B_2^{-} B_2^{+}\hat{B}_{k+4} -\frac{3 (k+4) (k+5) (13 k-25) }{8 (k+3)}  B_4^{-} B_2^{+}\hat{B}_{k+6} \nonumber\\
   & - \frac{\left(32 k^4+377 k^3+1612 k^2+2647 k+1596\right)  }{8 (k+3)} B_5^{-} B_3^{+}\hat{B}_{k+6} \nonumber\\
   & -\frac{3 (k+4) (k+5) (13 k-25) }{8 (k+3)} B_2^{-} B_4^{+}\hat{B}_{k+6} - \frac{\left(121 k^3+1068 k^2+2327 k+1596\right)} {8 (k+3)}  B_3^{-} B_5^{+}\hat{B}_{k+6} \nonumber\\
   & -\frac{\left(128 k^4+1787 k^3+8260 k^2+14101 k+8148\right) }{16 (k+3)}  B_4^{-} B_4^{+}\hat{B}_{k+6}\nonumber\\
   & - \frac{3 \left(101 k^3+1340 k^2+9931 k+21612\right)}{224 (k+3)} \left( B_6^{-} B_4^{+} + B_4^{-} B_6^{+}\right) \hat{B}_{k+6} \nonumber\\
   & -\frac{24 (k-1) k  }{(k+3)^2}B_2^{-} B_3^{+}\hat{B}_{k+3}-\frac{48 (k-1) k  }{(k+3)^2}B_2^{-} B_5^{+}\hat{B}_{k+3}-\frac{48 (k-1) k }{(k+2) (k+3)^2}B_3^{-} B_5^{+} \hat{B}_{k+2} \nonumber  \\
   & -\frac{96 k (k+1)  }{(k+3) (k+4)}B_3^{-} B_5^{+}\hat{B}_{k+4} -\frac{144 k (k+1)  }{(k+3)^2 (k+4)}B_4^{-} B_5^{+}\hat{B}_{k+3} \nonumber\\
   &  - \frac{\left(315 k^4+3744 k^3+16821 k^2+28712 k+23120\right) }{32 (k+3) (k+4)} B_3^{-} B_3^{+}\hat{B}_{k+4} \nonumber\\
   & -\frac{3 \left(101 k^3+1340 k^2+9931 k+21612\right)  }{56 (k+3) (k+5)}\left(B_6^{-} B_5^{+}+B_5^{-} B_6^{+}\right)\hat{B}_{k+5} \nonumber\\
   & -\frac{\left(3452 k^5+44021 k^4+244948 k^3+969163 k^2+3438360 k+4898736\right)   }{1120 (k+3) (k+4) (k+5)}B_4^{-}B_4^{+}\hat{B}_{k+4}\;.
   \nonumber
\end{align}
}
}

\end{adjustwidth}

\section{Conventions for $D$-functions}
\label{app:D-fns-conventions}

In this Appendix we record our conventions for $D$-functions, $\bar{D}$-functions and $\hat{D}$-functions.

In $d$ boundary dimensions, the $D$-functions $D_{\Delta_1 \Delta_2 \Delta_3 \Delta_4}(x_1, x_2, x_3, x_4)$ are defined as follows~\cite{DHoker:1999kzh},
\begin{align}
    D_{\Delta_1 \Delta_2 \Delta_3 \Delta_4}(x_1, x_2, x_3, x_4) &\,=\, \int d^{d+1} \boldsymbol{w} \sqrt{\bar{g}}\, \prod_{i=1}^4 K_{\Delta_i}(\boldsymbol{w}| \vec{x}) \\
    &\,=\, \Gamma\left(\frac{\hat{\Delta}-d}{2}\right) \frac{\pi^{d/2}}{2} \int_0^{\infty} \prod_{i=1}^4 \left[dt_i\frac{t_i^{\Delta_i-1}}{\Gamma(\Delta_i)}\right]\frac{1}{T^{\hat{\Delta}/2}} e^{-\sum_{i,j=1}^4 x_{ij}^2 \frac{t_i t_j}{2 T} }\,,  
    \nonumber
\end{align}
where $x_{ij}\equiv x_i-x_j\,,\,$ $x_{ij}^2\equiv |x_{ij}|^2$, $\,T\equiv\sum_i t_i\,$, and $\hat{\Delta}\equiv\sum_i \Delta_{i}\,$.

Recalling our convention for the conformal cross-ratios $U$ and $V$ given in the main text in Eq.~\eqref{eq:U-V-def}, we introduce the 
$\bar{D}$-functions $\bar{D}_{\Delta_1 \Delta_2 \Delta_3 \Delta_4}(U,V)$ via 
\be
\label{eq:Dbar-def}
  D_{\Delta_1 \Delta_2 \Delta_3 \Delta_4}(x_1, x_2, x_3, x_4) \,=\, \frac{\pi^{d/2}\Gamma\left(\frac{\hat{\Delta}-d}{2}\right)}{2\prod_{i=1}^4\Gamma(\Delta_i)} \frac{x_{14}^{\hat{\Delta}-2\Delta_1-2\Delta_4} x_{34}^{\hat{\Delta}-2\Delta_3-2\Delta_4}}{x_{13}^{\hat{\Delta}-2\Delta_4} x_{24}^{2\Delta_2}} \bar{D}_{\Delta_1 \Delta_2 \Delta_3 \Delta_4}(U,V)\,.  
\ee
It is common to parametrize the cross-ratios with a complex variable $z$, which we do as follows, 
\be
\label{eq:U-V}
    U = (1-z)(1-\bar{z})\,,\qquad V = z\bar{z}\;.
\ee
Given this, for $d=4$, we define the  
$\hat{D}$-functions as follows,
\be
\begin{split}
\label{eq:d-hat-defn}
     \hat{D}_{\Delta_1 \Delta_2 \Delta_3 \Delta_4}(z,\bar{z}) &\,=\, \lim_{|\vec{x}_2|\to\infty} |\vec{x}_2|^{2\Delta_2} D_{\Delta_1 \Delta_2 \Delta_3 \Delta_4}(\vec{x_1}=0,\vec{x_2},\vec{x_3}=\vec{n},\vec{x}) \\
     &\,=\, \frac{\pi^2\, \Gamma\left(\frac{\hat{\Delta}}{2}-2\right)}{2\prod_{i=1}^4 \Gamma(\Delta_i)} |z|^{\hat{\Delta}-2\Delta_1-2\Delta_4} |1-z|^{\hat{\Delta}-2\Delta_3-2\Delta_4} \bar{D}_{\Delta_1 \Delta_2 \Delta_3 \Delta_4}(z,\bar{z})\,. 
\end{split}
\ee


\vspace{8mm}

\section{Supergravity correlators in terms of D-functions}
\label{app:sugra-corrs}

In this appendix we record the outcome of the supergravity calculations of Section~\ref{sec:probe-calcs}, as described in the main text below Eq.\;\eqref{eq:g-dyna}. For Sequence 1, we obtain:
\begin{adjustwidth}{-4mm}{-4mm} 

\vspace{5mm}

\scalebox{0.96}{%
\parbox{\linewidth}{%
\begin{align}
     \cG_{3,3,k+4,k+4} \;=&\; \frac{1}{4} (k+3) (k+4) \hat{D}_{1,1,k+4,k+4}+\frac{3}{4} (k+4) (3 k-4) \hat{D}_{1,2,k+5,k+4}+\frac{1}{4} (k-12) (k+4) \hat{D}_{2,1,k+5,k+4}\nonumber\\
     &+\frac{24 k \left(k^2+k-2\right) \hat{D}_{1,4,k+3,k+4}}{(k+3)^2}+\frac{\left(105 k^3+614 k^2+477 k-1964\right) \hat{D}_{2,3,k+5,k+4}}{8 (k+3)}\nonumber\\
     &+\frac{24 (k-1) k \hat{D}_{2,4,k+2,k+4}}{(k+3)^2}+\frac{24 k (k+1) \hat{D}_{2,4,k+4,k+4}}{k+3}-\frac{24 k (k+1) \hat{D}_{2,5,k+5,k+4}}{k+3}\nonumber\\
     &+\frac{\left(57 k^3+406 k^2+413 k-1964\right) \hat{D}_{3,2,k+5,k+4}}{8 (k+3)}+\frac{48 k (k+1) \hat{D}_{3,4,k+3,k+4}}{(k+3)^2}\nonumber\\
     &+\frac{\left(799 k^4+10368 k^3+46913 k^2+85368 k+66000\right) \hat{D}_{3,4,k+5,k+4}}{32 (k+3) (k+5)}\nonumber\\
     &+\frac{\left(128 k^5+2591 k^4+19072 k^3+63553 k^2+94968 k+66000\right) \hat{D}_{4,3,k+5,k+4}}{32 (k+3) (k+5)}\nonumber\\
     &+\frac{\left(4967 k^4+36269 k^3+115669 k^2+1200379 k+2690076\right) \hat{D}_{4,5,k+5,k+4}}{1120 (k+3) (k+5)}\nonumber\\
     &+4 \left(k^3+9 k^2+20 k+12\right)\left( \hat{D}_{4,5,k+7,k+4}+\hat{D}_{5,4,k+7,k+4}\right) \nonumber\\
     &+\frac{\left(-8473 k^4-71251 k^3-112811 k^2+1065979 k+2690076\right) \hat{D}_{5,4,k+5,k+4}}{1120 (k+3) (k+5)}\nonumber\\
     &+\frac{\left(863 k^4+14096 k^3+68641 k^2+202696 k+272784\right) \hat{D}_{5,5,k+4,k+4}}{70 (k+3) (k+4) (k+5)} \\
     &+\frac{\left(1937 k^3+9420 k^2-5153 k-51396\right) \hat{D}_{5,5,k+6,k+4}}{140 (k+3)}-\frac{24 (k-1) k \hat{D}_{1,5,k+4,k+4}}{k+3}\nonumber\\
     &-\frac{(k-5) (k+4) (k+5) \left(\hat{D}_{1,3,k+6,k+4}+\hat{D}_{3,1,k+6,k+4}\right)}{2 (k+3)}-\frac{24 (k-1) k \hat{D}_{2,3,k+3,k+4}}{(k+3)^2}\nonumber\\
     &-\frac{(k+4) (k+5) (3 k+1) \hat{D}_{2,2,k+6,k+4}}{2 (k+3)}-\frac{(k+4) (k+5) (21 k-17) \hat{D}_{3,3,k+6,k+4}}{2 (k+3)}\nonumber\\
     &-\frac{(k+4) \left(9 k^2+32 k-81\right) \hat{D}_{2,2,k+4,k+4}}{4 (k+3)}-\frac{144 k (k+1) \hat{D}_{4,5,k+3,k+4}}{(k+3)^2 (k+4)}\nonumber\\
     &-\frac{3 (k+4) (k+5) (13 k-25) \left(\hat{D}_{2,4,k+6,k+4}+\hat{D}_{4,2,k+6,k+4}\right)}{8 (k+3)}\nonumber\\ &-\frac{\left(121 k^3+1068 k^2+2327 k+1596\right) \hat{D}_{3,5,k+6,k+4}}{8 (k+3)}-\frac{48 (k-1) k \hat{D}_{2,5,k+3,k+4}}{(k+3)^2}\nonumber\\
     &-\frac{\left(32 k^4+377 k^3+1612 k^2+2647 k+1596\right) \hat{D}_{5,3,k+6,k+4}}{8 (k+3)}-\frac{48 (k-1) k \hat{D}_{3,5,k+2,k+4}}{(k+2) (k+3)^2}\nonumber\\
     &-\frac{\left(128 k^4+1787 k^3+8260 k^2+14101 k+8148\right) \hat{D}_{4,4,k+6,k+4}}{16 (k+3)}-\frac{96 k (k+1) \hat{D}_{3,5,k+4,k+4}}{(k+3) (k+4)}\nonumber\\
     &-\frac{3 \left(101 k^3+1340 k^2+9931 k+21612\right) \left(\hat{D}_{4,6,k+6,k+4}+\hat{D}_{6,4,k+6,k+4}\right)}{224 (k+3)}\nonumber\\
     &-\frac{3 \left(101 k^3+1340 k^2+9931 k+21612\right) \left(\hat{D}_{5,6,k+5,k+4}+\hat{D}_{6,5,k+5,k+4}\right)}{56 (k+3) (k+5)}\nonumber\\
     &-\frac{\left(315 k^4+3744 k^3+16821 k^2+28712 k+23120\right) \hat{D}_{3,3,k+4,k+4}}{32 (k+3) (k+4)}\nonumber\\
    &-\frac{\left(3452 k^5+44021 k^4+244948 k^3+969163 k^2+3438360 k+4898736\right) \hat{D}_{4,4,k+4,k+4}}{1120 (k+3) (k+4) (k+5)}
    \nonumber\;.
\end{align}
}%
}

\end{adjustwidth}


For Sequence 2, we obtain:
\begin{align}
     \cG_{2,3,k+4,k+5} \;=&\; -\frac{18}{35} \left(17 k^2+113 k+140\right) \hat{D}_{2,5,k+6,k+5}+4 (k+3) \left(k^2+8 k+12\right) \hat{D}_{3,5,k+7,k+5} \nonumber\\
     &+\frac{4}{35} \left(109 k^2+776 k+1470\right) \hat{D}_{3,6,k+6,k+5}+3 (k+3) \left(k^2+8 k+12\right) \hat{D}_{4,4,k+7,k+5} \nonumber\\
     &+\frac{24}{175} \left(127 k^2+698 k+1050\right) \hat{D}_{4,5,k+6,k+5}+\frac{12 \left(109 k^2+776 k+1470\right) \hat{D}_{4,6,k+5,k+5}}{35 (k+5)} \nonumber\\
     &-\frac{36}{35} \left(2 k^2+17 k+28\right) \hat{D}_{5,4,k+6,k+5}-\frac{144 \left(2 k^2+17 k+28\right) \hat{D}_{5,5,k+5,k+5}}{35 (k+5)} \nonumber\\
     &-\frac{48 k \hat{D}_{3,5,k+3,k+5}}{(k+3)(k+4)}+\frac{6 \left(47 k^3+510 k^2+1858 k+2100\right) \hat{D}_{2,4,k+5,k+5}}{35 (k+5)} \nonumber\\
     &+\left(-7 k^3-\frac{2643 k^2}{35}-\frac{8392 k}{35}-216\right) \hat{D}_{3,4,k+6,k+5} \nonumber\\
     & -\frac{3}{7} \left(7 k^3+54 k^2+109 k+28\right) \hat{D}_{4,3,k+6,k+5}  \nonumber\\
     &-\frac{\left(5 k^3+43 k^2+262 k+200\right) \hat{D}_{2,3,k+4,k+5}}{5 (k+4)} \nonumber\\
     &-\frac{12 \left(251 k^3+3097 k^2+13312 k+18200\right) \hat{D}_{3,5,k+5,k+5}}{175 (k+5)}\\
     &-\frac{6 \left(361 k^3+2280 k^2+3299 k-1260\right) \hat{D}_{4,4,k+5,k+5}}{175 (k+5)} \nonumber\\
     &-\frac{12 \left(317 k^3+2370 k^2+7243 k+7140\right) \hat{D}_{4,5,k+4,k+5}}{175 (k+4) (k+5)} \nonumber\\
     &+\frac{\left(105 k^4+1475 k^3+6768 k^2+11252 k+3920\right) \hat{D}_{3,3,k+5,k+5}}{35 (k+5)} \nonumber\\
     &+\frac{3 \left(317 k^4+4447 k^3+22268 k^2+52068 k+45640\right) \hat{D}_{3,4,k+4,k+5}}{175 (k+4) (k+5)} \nonumber\\
     &+\frac{24 k (k+3) \hat{D}_{1,4,k+4,k+5}}{k+4}-\frac{12}{5} (k-1) (k+5) \hat{D}_{1,4,k+6,k+5} -24 k \hat{D}_{1,5,k+5,k+5} \nonumber\\
     &+\frac{2}{5} (k-1) k \hat{D}_{2,2,k+5,k+5} -\frac{8}{5} (k+5) (2 k+3) \hat{D}_{2,3,k+6,k+5}+\frac{24 k \hat{D}_{2,4,k+3,k+5}}{k+4} \nonumber\\
     &-\frac{2}{5} (k-6) (k+5) \hat{D}_{3,2,k+6,k+5}-\frac{48 k \hat{D}_{2,5,k+4,k+5}}{k+4}+\frac{4}{5} k (7 k+23) \hat{D}_{1,3,k+5,k+5} \nonumber\;.
\end{align}

\end{appendix}


\newpage

\begin{adjustwidth}{-3mm}{-3mm} 
\bibliographystyle{utphys}      
\bibliography{microstates}       

\end{adjustwidth}


\end{document}